\documentclass[pdflatex,iicol]{sn-jnl}
\usepackage{comment}
\usepackage{amsmath}
\usepackage{bm}
\usepackage{amsfonts}
\usepackage{tikz}
\usetikzlibrary{positioning}
\usepackage{cite}
\usepackage{enumerate}
\usepackage{graphicx}
\usepackage{hyperref}
\usepackage{multicol}
\usepackage{multirow}
\usepackage[normalem]{ulem}
\usepackage{xcolor}
\usepackage{subfigure}

\newcommand{\norm}[1]{\left\lVert#1\right\rVert}
\newcommand{\purplesout}[1]{} 																											  

\begin{document}

\title{Design and Analysis of Binaural Signal Matching with Arbitrary Microphone Arrays and Listener Head Rotations}

\author[1]{\fnm{Lior} \sur{Madmoni}}\email{liomad@gmail.com}
\author[2]{\fnm{Zamir} \sur{Ben-Hur}}\email{zamirbh@fb.com}
\author[2]{\fnm{Jacob} \sur{Donley}}\email{jdonley@fb.com}
\author[2]{\fnm{Vladimir} \sur{Tourbabin}}\email{vtourbabin@fb.com}
\author[1]{\fnm{Boaz} \sur{Rafaely}}\email{br@bgu.ac.il}

\affil[1]{\orgdiv{School of Electrical and Computer Engineering}, \orgname{Ben-Gurion University of the Negev}, \orgaddress{\city{Beer-Sheva}, \postcode{84105}, \country{Israel}}}

\affil[2]{\orgdiv{Reality Labs @ Meta}, \orgaddress{\city{Redmond}, \state{WA}, \country{USA}}}

\abstract{
    Binaural reproduction is rapidly becoming a topic of great interest in the research community, especially with the surge of new and popular devices, such as virtual reality headsets, smart glasses, and head-tracked headphones. 
    In order to immerse the listener in a virtual or remote environment with such devices, it is essential to generate realistic and accurate binaural signals. 
    This is challenging, especially since the microphone arrays mounted on these devices are typically composed of an arbitrarily-arranged small number of microphones, which impedes the use of standard audio formats like Ambisonics, and provides limited spatial resolution. 
    The binaural signal matching (BSM) method was developed recently to overcome these challenges. While it produced binaural signals with low error using relatively simple arrays, its performance degraded significantly when head rotation was introduced. 
    This paper aims to develop the BSM method further and overcome its limitations. For this purpose, the method is first analyzed in detail, and a design framework that guarantees accurate binaural reproduction for relatively complex acoustic environments is presented. Next, it is shown that the BSM accuracy may significantly degrade at high frequencies, and thus, a perceptually motivated extension to the method is proposed, based on a magnitude least-squares (MagLS) formulation. These insights and developments are then analyzed with the help of an extensive simulation study of a simple six-microphone semi-circular array. It is further shown that the BSM-MagLS method can be very useful in compensating for head rotations with this array. Finally, a listening experiment is conducted with a four-microphone array on a pair of glasses in a reverberant speech environment and including head rotations, where it is shown that BSM-MagLS can indeed produce binaural signals with a high perceived quality.
}

\maketitle

\section{Introduction}
\label{sec:intro}
\begin{figure}
	\centering
	\begin{tikzpicture}
		\node (img)  {\includegraphics[width=0.99\columnwidth,trim={0cm 0cm 0cm 0cm},clip]{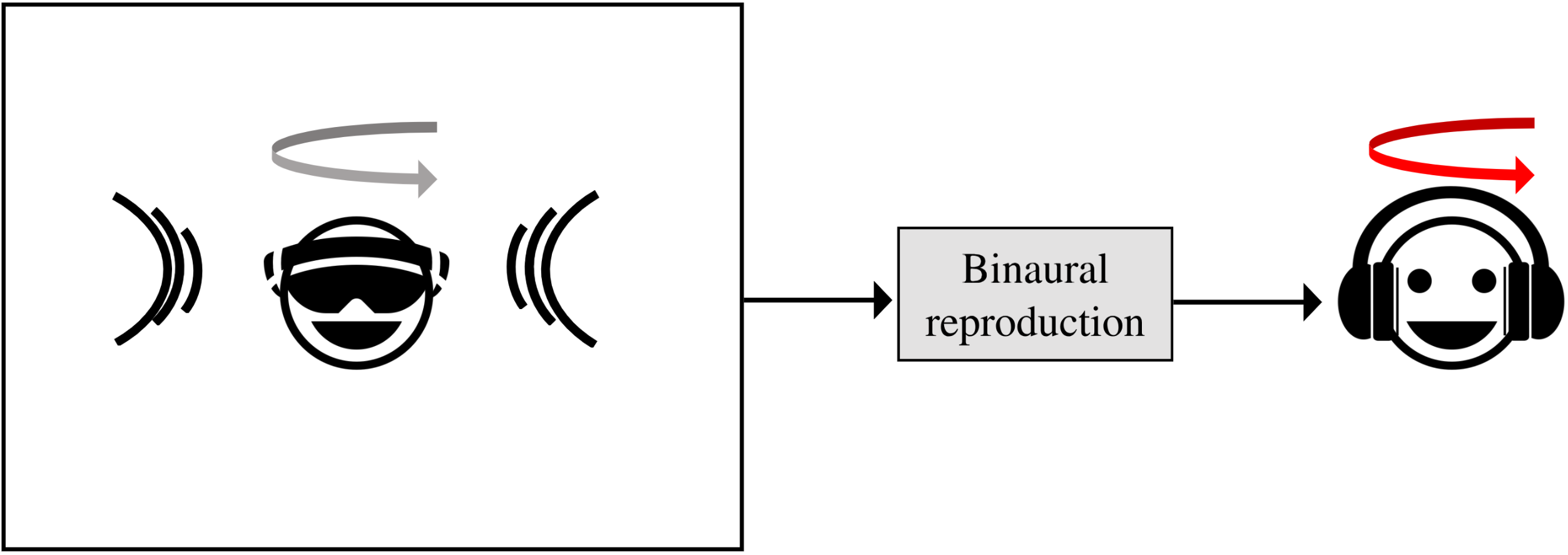}};
	\end{tikzpicture}
	\caption{An illustration of headphones binaural reproduction. The person on the left records the acoustic environment using a wearable array. The microphone signals are then processed and transmitted to a remote listener where binaural reproduction is performed with headphones. Both the person with the wearable array and the listener are free to move their heads, denoted by the gray and red curved arrows, respectively.}
	\label{fig:acoustic_environment}	
\end{figure}

Binaural reproduction is an ongoing research topic with an increasing number of applications for augmented and virtual reality, teleconferencing and hearing aids \cite{rafaely2022spatial}. To binaurally reproduce an acoustic scene, the sound field and the head-related transfer functions (HRTFs) are required. In real-life acoustic scenes, these can be captured simultaneously using microphones that are positioned in the ears of a listener or an anatomically equivalent dummy. However, this approach limits the separation of the sound field and HRTF components, impeding the reproduction of the sound field for different head orientations, or using a different HRTF for personalization. To overcome this limitation, an array of microphones is typically required. 
Therefore, binaural reproduction with microphone arrays has become a topic of interest recently. 
An illustration of headphone binaural reproduction is presented in Fig.~\ref{fig:acoustic_environment}, showing a person recording sound from the environment using a wearable array on the left. The recorded signals are then processed to produce binaural signals. These are then played back over headphones to a remote person, shown on the right.

A common approach to decouple the sound field from the effect of the head is to use higher order Ambisonics (HOA) signals for binaural reproduction \cite{HOA_1995analysis,HOA_2003further}. Systems that capture and reproduce HOA were proposed in \cite{HOA_2006_4th_order} using spatial coding, in \cite{rafaely2010interaural} using spherical microphone arrays and in \cite{poletti2005three} for reproduction via loudspeaker arrays. 
While HOA is a well studied format, it often requires arrays with high directional resolution and a specific configuration, such as spherical arrays. 
For non-spherical arrays, \cite{ahrens2021equatorial} proposed a solution for elevation-invariant sound-fields with fully-circular arrays. It was later extended in \cite{ahrens2022circumferential} to more general array-geometries of circumferential contours around non-spherical bodies using a numerical least-squares fitting. However, the study included a relatively large array, consisting of 18 microphones. 
Another work \cite{mccormack2022parametric_3plus1} studied Ambisonics reproduction with smaller arrays, but it is limited to a dedicated microphone array with a ``three-plus-one'' configuration. 
To summarize, Ambisonics-based reproduction is currently missing a clear framework for smaller arrays and arbitrary microphone positions, such as mobile, handheld or wearable arrays, that are often desired for many applications.

In order to overcome these limitations, one popular approach is to use parametric methods for binaural reproduction, such as \cite{HARPEX, COMPASS_1,pulkki2018parametric}. For example, \cite{fernandez2022enhancing} used an eight-microphone array mounted on a pair of glasses and proposed an enhancement stage using covariance matching, while \cite{mccormack2022parametric_3plus1} and \cite{mccormack2019parametric} proposed Ambisonics reproduction from non-spherical arrays. However, the resulting quality depends on the estimation accuracy of the model parameters, such as the direction-of-arrival, diffuseness of the sound-field, and on the sparsity assumption of the sound sources in the time-frequency domain. In addition, calculating these parameters may increase the computational complexity, compared to a signal-independent approach. Furthermore, this may increase the complexity of a design framework for binaural reproduction with general arrays while no guarantee on performance is presented, and thus, a non-parametric, signal-independent approach is suggested here.

A signal-independent approach that may be more suitable to arbitrary array geometries is beamforming-based binaural reproduction (BFBR) \cite{BFBR_1, BFBR_2}. However, the design parameters of BFBR should be set carefully in order to approximate the binaural signals closely. These parameters include the beamformer type, the steering directions and their number, and relative attenuation factors for each beamformer. Recent works have proposed various designs and studied the quality of the resulting binaural signals. 
For instance, \cite{BFBR_1} studied plane-wave decomposition beamformers steered towards the available HRTF directions in the spherical harmonics (SH) domain. This design was also studied in \cite{BFBR_2} for headphone reproduction and with steering directions that are chosen according to the main-lobe width of the beam-pattern \cite{van2004optimum}. 
The quality of such designs was further studied in \cite{BFBR_3} and \cite{BFBR_4}. 
While providing some useful insights, these works only studied spherical arrays, and did not address the incorporation of the design framework in other array geometries.

Other works that studied BFBR designs for other array geometries include \cite{BFBR_5_Calamaia}; the authors applied BFBR with minimum variance distortionless response (MVDR) beamformers and a microphone array mounted on a helmet with the aim of preserving binaural cues for azimuth localization. Other beamformer types were also studied and compared, such as the plane-wave decomposition and delay-and-sum beamformers in \cite{spors2012comparison}, and the MVDR and minimum mean-squared error (MMSE) beamformers in \cite{zhao20123d}. These works highlighted the advantages of using specific beamformer types, but were not extended to a more general design framework. However, such a framework was recently developed in \cite{Itay_BFBR} for spherical arrays, fully describing how to design a BFBR system that operates with HOA reproduction. In addition, it includes a guideline for choosing the number of maximum directivity beamformers when using arbitrary arrays. However, a theoretical framework for the design of the remaining parameters was not given, impeding the use of BFBR with these arrays. 
Moreover, an inherent limitation of these works is that they do not directly minimize the error of the desired binaural signals, and thus, ensuring the quality of the reproduction remains a challenge.

A third approach for binaural reproduction, which is also flexible in the array geometry, is binaural signals matching (BSM). 
This usually refers to the estimation of binaural signals while minimizing the mean-squared error (MSE) by matching the array steering vectors to the HRTFs using a linear formulation. 
Examples for such methods include \cite{VAH_2011}, which optimized the microphone positions in a planar array geometry. It was later extended to include HRTF smoothing for the purpose of producing perceptually accurate binaural signals with a smaller number of microphones \cite{VAH_2012}, and to include different regularization techniques for the MSE minimization \cite{VAH_2014,VAH_2016}. Furthermore, the method was evaluated with a perceptual study that showed it can be used with individual HRTFs and outperform generic HRTFs, but mainly for directions which were directly optimized in the derivations of the filter coefficients \cite{VAH_2017}. While these works proved that BSM can produce high quality binaural signals, they only studied relatively complex microphone arrays, consisting of 24 high-quality sensors in a planar geometry. This limits the use of BSM with more general array geometries.
Another work which studied a least-squares-based reproduction and integrated a magnitude least-squares (MagLS) solution is described in \cite{MagLS_schorkhuber2018binaural}, but the study was limited to Ambisonics signals. This work was extended in \cite{e2e_magls} by integrating the array model into the MagLS objective function, thus providing an end-to-end MagLS (eMagLS) binaural renderer. 
While the methods described in \cite{e2e_magls} can be used with any array geometry, the developments and evaluations are limited to spherical and equatorial microphone arrays. Implementing these arrays in practice can be relatively complex, requiring specific designs that may be impractical for arrays mounted on small devices. Thus, later works studied methods with smaller arrays having only a few microphones, such as head-worn arrays \cite{A_stahl2023perceptual,B_mccormack2023six,C_deppisch2024blind}. These works highlight the superiority of matching to binaural signals, also denoted as eMagLs, for such arrays, compared to other popular non-parametric methods, such as Ambisonics rendering and BFBR. However, without a rigorous theoretical analysis, the conditions under which matching to binaural signals guarantees accurate binaural reproduction remain unclear, limiting its wider applicability and optimization.


	
In this paper, a theoretical framework for the design of BSM systems, including eMagLS, is developed. The framework provides a theoretical basis for analyzing the performance of BSM reproduction with any array geometry.
Following the theoretical analysis of BSM, the method is numerically studied using MSE and perceptually-motivated measures with a semi-circular array. 
Finally, a listening experiment that includes a four-microphone array that is mounted on a pair of glasses is presented to validate the theoretical results.

The contributions of this paper are as follows.
\begin{enumerate}
	\item A design framework for binaural reproduction with arbitrary microphone arrays is developed, including a mathematical development showing that it is a generalization of a beamforming-based design.
	\item Explicit conditions for accurate binaural reproduction with BSM are developed, extending the method to arbitrary sound fields.
	\item A comprehensive simulation study and a listening experiment validate the accuracy and quality of BSM with small microphone arrays. The study includes head rotations and MagLS, which further motivates the incorporation of BSM in wearable and mobile devices.
\end{enumerate}

\section{Background}
\label{sec:background}
\begin{figure}
    \centering
    \begin{minipage}{1\columnwidth}	
        \centering
        \begin{tikzpicture}
            \node (img)  {\includegraphics[width=0.99\columnwidth,trim={0cm 0cm 0cm 0cm},clip]{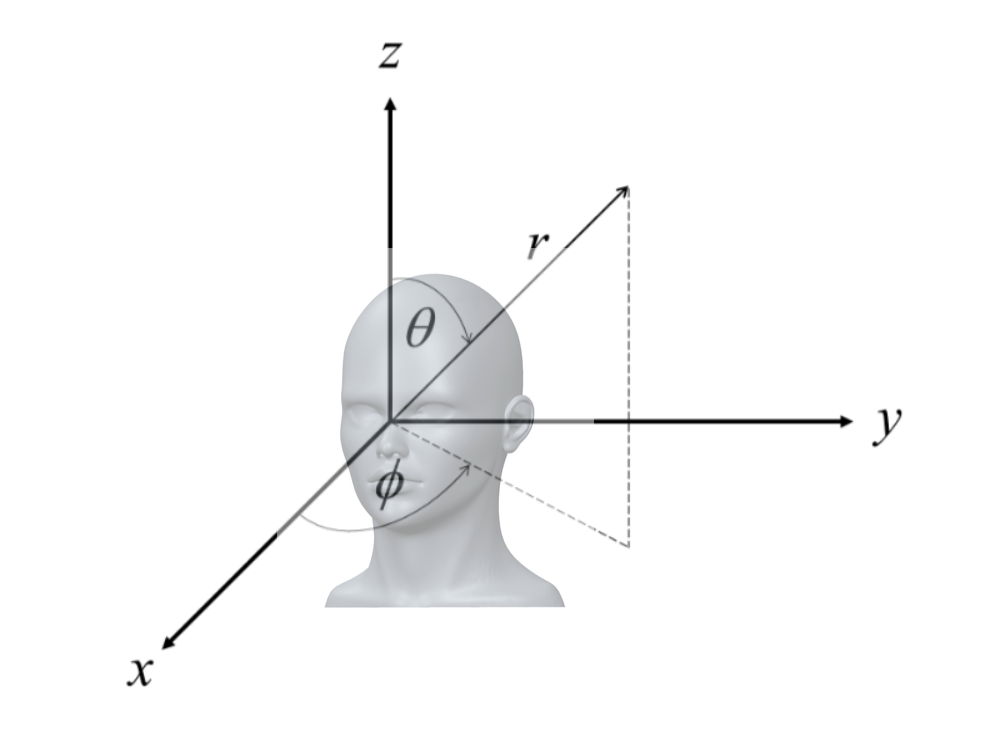}};
        \end{tikzpicture}	
    \end{minipage}	
    \begin{minipage}{1\columnwidth}
        \centering
        \caption{An illustration of the spherical coordinate system and the reference head orientation that is used in this study.}
            \label{fig:coordinates_w_head}
    \end{minipage}
\end{figure}

This section presented array processing models and binaural reproduction methods. Throughout the paper, a spherical coordinates system will be used, denoted $(r,\theta,\phi)$, where $r$ is the distance from the origin, $\theta$ is the polar angle measured downward from the Cartesian $z$ axis to the $xy$ plane, and $\phi$ is the azimuthal angle measured from the positive $x$ axis towards the positive $y$ axis. 
This coordinate system including the corresponding reference head orientation that will be used in this study is illustrated in Fig.~\ref{fig:coordinates_w_head}.

\subsection{Microphone Array Measurement Model}
Assume that an $M$-element microphone array, centered at the origin, is used to capture a sound field that is comprised of $Q$ far-field sources that are carrying the signals $\{s_q(k)\}_{q=1}^{Q}$ with the corresponding directions-of-arrival (DOAs) $\{(\theta_q, \phi_q)\}_{q=1}^{Q}$. Here, $k=\frac{2\pi}{\lambda}$ is the wave-number and $\lambda$ is the wave-length. Then, the pressure that is measured by the array can be described by the following narrow-band model \cite{van2004optimum}: 
\begin{align}
\mathbf{x}(k) = \mathbf{V}(k) \mathbf{s}(k) + \mathbf{n}(k),
\label{eq:array_meas_VanTrees}
\end{align}
where $\mathbf{x}(k) = \begin{bmatrix}
x_1(k),  x_2(k),  \ldots ,  x_M(k)
\end{bmatrix}^T$ 
is a vector of length $M$ containing the microphone signals, $ \mathbf{V}(k) = \begin{bmatrix}
\mathbf{v}_1(k),  \mathbf{v}_2(k),  \ldots , \mathbf{v}_Q(k)
\end{bmatrix}$ 
is an $M\times Q$ matrix with its $q$-th column containing the steering vector of the $q$-th source for $q=1,2,\ldots,Q$, $\mathbf{s}(k)=\begin{bmatrix}
s_1(k),  s_2(k),  \ldots, s_Q(k)
\end{bmatrix}^T$ is a vector of length $Q$ containing the source signals, $\mathbf{n}(k) = \begin{bmatrix}
n_1(k),  n_2(k),  \ldots ,  n_M(k)
\end{bmatrix}^T$ is an additive noise vector of length $M$, and $(\cdot)^T $ is the transpose operator. The set of steering vectors can be described as
\begin{align}
    \mathbf{v}_q(k) =\begin{bmatrix}
    v(k, \mathbf{d}_1 ; \theta_q, \phi_q)\\
        v(k, \mathbf{d}_2 ; \theta_q, \phi_q)\\
        \vdots\\
        v(k, \mathbf{d}_M ; \theta_q, \phi_q)
    \end{bmatrix}, \quad\forall q=1,...,Q,
\label{eq:ATF}
\end{align}
where $\mathbf{d}_m$ is the Cartesian coordinates of the $m$-th microphone in the array and $v(k, \mathbf{d}_m ; \theta_q, \phi_q)$ is the transfer function between a far-field source with a DOA of $(\theta_q, \phi_q)$ and $\mathbf{d}_m $ for $m=1,2,\ldots,M$. These steering vectors can be calculated analytically for various array types \cite{van2004optimum,rafaely2015fundamentals}, numerically, or measured.

\subsection{Binaural Signal Representation using HRTFs}
Assume that a listener is surrounded by a sound field that can be described by the plane-wave density (PWD) function $a(k,\theta,\phi)$. Then, the sound pressure at the listener's ears can be described by \cite{menzies2007nearfield}:
\begin{align}
	p^{l,r}(k) = \int_{0}^{2\pi} \int_{0}^{\pi} a(k,\theta,\phi) h^{l,r}(k,\theta,\phi) \sin\theta d\theta d\phi,
	\label{eq:binau_repr_space}
\end{align}
where $p^{l,r}(k)$ are the sound pressure values and $h^{l,r}(k,\theta,\phi)$ are the HRTFs of the left and right ears, denoted by $(\cdot)^{l}$ and $(\cdot)^{r}$, respectively. 
When the sound field is comprised of $Q$ far-field sources, the binaural signals in (\ref{eq:binau_repr_space}) can be further reduced to: 
\begin{align}
	p^{l,r}(k) = \sum_{q=1}^{Q} s_q(k) h^{l,r}(k,\theta_q, \phi_q) = [\mathbf{h}^{l,r}(k)]^T \mathbf{s}(k) , 
	\label{eq:binau_repr_space_L_PWs}
\end{align}
where 
\begin{align*}
    \mathbf{h}^{l,r}(k) =
    \begin{bmatrix}
        h^{l,r}(k, \theta_1 , \phi_1)\\ h^{l,r}(k, \theta_2 , \phi_2)\\ \vdots\\ h^{l,r}(k, \theta_Q , \phi_Q)
    \end{bmatrix}
\end{align*}
is a vector of length $Q$ containing the HRTFs corresponding to the $Q$ directions of the sources. 
These signals can be transformed to the time domain using the inverse discrete Fourier transform (IDFT) applied to (\ref{eq:binau_repr_space_L_PWs}) up to the Nyquist frequency. Then, they can be played back over headphones for binaural reproduction. 

\subsection{Beamforming-based Binaural Reproduction}
\label{subsec:BFBR}
A method for binaural reproduction based on beamforming is presented next. In the first stage of this method, the PWD is estimated at a pre-defined set of $D$ directions $\{(\theta_d, \phi_d)\}_{d=1}^D$. This is performed by spatially filtering the signals in (\ref{eq:array_meas_VanTrees}) using beamformers with $D$ look-directions, as in the following \cite{van2004optimum}:
\begin{align}
y(k, \theta_d, \phi_d) = \mathbf{w}^H_d(k) \mathbf{x}(k),  \quad \forall d=1,2,\ldots, D,
\label{eq:beamforming}
\end{align} 
where the beamformer output $y(k, \theta_d, \phi_d)$ constitutes the estimated PWD,
\begin{align}
	\mathbf{w}_d(k) = \begin{bmatrix}
		w_1(k, \theta_d, \phi_d)\\ w_2(k, \theta_d, \phi_d)\\ \vdots\\ w_M(k, \theta_d, \phi_d)
	\end{bmatrix}
	\label{eq:w_d}	
\end{align}
is a vector of length $M$ holding the beamformer weights for $d=1,2,\ldots, D$, and $(\cdot)^H$ is the Hermitian operator. Similarly to (\ref{eq:binau_repr_space_L_PWs}), the beamformer outputs are then scaled, multiplied by HRTFs from the corresponding look-directions and summed to produce an estimated binaural signal \cite{Itay_BFBR}:
\begin{align}
\hat{p}_{\text{BFBR}}^{l,r}(k) = \sum_{d=1}^{D} \alpha_d y(k, \theta_d, \phi_d) h^{l,r}(k, \theta_d, \phi_d),
\label{eq:BFBR}
\end{align}
where $\{\alpha_d\}_{d=1}^D$ are the scaling factors. 
In BFBR design, the beamformer weights, the set of look-directions and the scaling factors are chosen with the aim of producing the desired binaural signals. The scaling factors provide additional control over the reproduced signal. For instance, certain look-directions that may be of high importance can be amplified by controlling the corresponding $\alpha_d$. In addition, with spherical arrays they can be set to produce binaural signals that are equivalent to HOA reproduction \cite{Itay_BFBR}.

\subsection{Binaural Reproduction using Ambisonics}
Ambisonics signals refer to the SH representation of the PWD function, denoted $a_{nm}(k)$ with SH order $n$ and degree $m$ \cite{rafaely2015fundamentals}. In practice, these are captured by an array with limited spatial resolution, such that the PWD function can be extracted accurately up to a maximal SH order of $N_a$ \cite{rafaely2015fundamentals}. The HRTFs are also typically measured or calculated up to a finite SH order, denoted $N_H$. 
Binaural reproduction can then be performed with the SH representation of (\ref{eq:binau_repr_space}) \cite{rafaely2010interaural}:
\begin{align}
	\hat{p}^{l,r}_{\text{HOA}}(k) = \sum_{n=0}^{N_p} \sum_{m=-n}^n \tilde{a}^*_{nm}(k) h_{nm}^{l,r}(k),
	\label{eq:binau_repr_SH_finite}
\end{align}
where $\tilde{a}_{nm}(k)$ and $h_{nm}^{l,r}(k)$ are the spherical Fourier transform (SFT) coefficients of the complex conjugate of $a(k,\theta,\phi)$ and of $h^{l,r}(k,\theta,\phi)$, respectively, $N_p = \min\{N_a, N_H\}$ and $(\cdot)^*$ is the complex conjugate operator. 
Equation (\ref{eq:binau_repr_SH_finite}) is using the Ambisonics signal, $a_{nm}(k)$, explicitly for binaural reproduction \cite{HOA_1995analysis}, and it is presented here since this formulation will be used in the listening experiment as a reference.

The Ambisonics format can also be formulated using a special design of BFBR. This formulation was recently described in \cite{Itay_BFBR} for spherical arrays. It was shown that the BFBR output in (\ref{eq:BFBR}) is equivalent to the HOA reproduction in (\ref{eq:binau_repr_SH_finite}) when maximum directivity beamformers are used with a set of look-directions $\{(\theta_d, \phi_d)\}_{d=1}^D$ that correspond to an aliasing-free sampling scheme on the sphere up to SH orders of $\max\{N_H, N_a\}$, and setting the scaling factors $\{\alpha_d\}_{d=1}^D$ according to the corresponding sampling weights. For these conditions to hold, the number of beamformers $D$ should be greater than or equal to the directivity factor of the maximum directivity beamformer \cite{van2004optimum,rafaely2015fundamentals}. It was also suggested in \cite{Itay_BFBR} to use the average directivity factor of the maximum directivity beamformer when using arbitrary array geometries. However, a full framework for these arrays has not yet been developed.

\section{Proposed Method for Binaural Signal Matching}
\label{sec:method}
This section describes the proposed BSM method for binaural reproduction with general arrays. This method produces time-invariant multiple-input-multiple-output (MIMO) filters in the frequency domain. 
While the derivations of BSM shown here were already presented in previous works \cite{VAH_2011,delikaris2015parametric,fernandez2022enhancing,e2e_magls}, they are repeated here for completeness and to highlight the specific design parameters of BSM that are analyzed in the following section. These parameters are crucial for generating accurate binaural signals. 
For brevity, the wave-number index, $k$, will be omitted.

\subsection{Formulation of the BSM Approach}
In the first stage of deriving the BSM method, the microphone signals are spatially filtered according to
\begin{align}
\hat{p}_{\text{BSM}}^{l,r} = [\mathbf{c}^{l,r}]^H \mathbf{x},
\label{eq:z_lr}
\end{align}
where $\hat{p}_{\text{BSM}}^{l,r} $ are the estimated binaural signals according to the BSM method, and $\mathbf{c}^{l,r}$ are vectors of length $M$ holding the filter coefficients for the left and rights ears that will be formulated below. Next, the following MSE between the binaural signals in (\ref{eq:binau_repr_space_L_PWs}) and the filtered microphone signals in (\ref{eq:z_lr}) is calculated for each ear separately:
\begin{align}
\epsilon^{l,r} = \mathbb{E}\begin{bmatrix}
|p^{l,r} - \hat{p}_{\text{BSM}}^{l,r}|^2
\end{bmatrix},
\label{eq:eps_bin_1}
\end{align}
where $\mathbb{E}\begin{bmatrix}
\cdot
\end{bmatrix}$ is the expectation operator. Assuming that the source signals, $\{s_q\}_{q=1}^{Q}$, are uncorrelated with the noise components, $\{n_m\}_{m=1}^{M}$, and substituting (\ref{eq:array_meas_VanTrees}), (\ref{eq:binau_repr_space_L_PWs}) and (\ref{eq:z_lr}) in (\ref{eq:eps_bin_1}), leads to the following MSE:
\begin{align}
\nonumber
\epsilon^{l,r} =  & \mathbb{E}\left[
|[\mathbf{h}^{l,r}]^T \mathbf{s} - [\mathbf{c}^{l,r}]^H \big(\mathbf{V} \mathbf{s} + \mathbf{n}\big)|^2
\right]
\\
\nonumber
= & \mathbb{E}\left[
|\big([\mathbf{h}^{l,r}]^T - [\mathbf{c}^{l,r}]^H \mathbf{V} \big) \mathbf{s} - [\mathbf{c}^{l,r}]^H  \mathbf{n}\big)|^2
\right]
\\
\nonumber
= & \mathbb{E}\left[
|\big([\mathbf{c}^{l,r}]^H\mathbf{V} - [\mathbf{h}^{l,r}]^T\big) \mathbf{s} + [\mathbf{c}^{l,r}]^H  \mathbf{n}\big)|^2
\right]
\\
= & 
\big([\mathbf{c}^{l,r}]^H \mathbf{V} - [\mathbf{h}^{l,r}]^T\big) \mathbf{R_s} \big([\mathbf{c}^{l,r}]^H \mathbf{V} - [\mathbf{h}^{l,r}]^T\big)^H \nonumber \\
& + [\mathbf{c}^{l,r}]^H \mathbf{R_n} [\mathbf{c}^{l,r}],
\label{eq:eps_bin_2}
\end{align}
where, in the fourth line of (\ref{eq:eps_bin_2}) we have used the assumption that the source and noise signals are uncorrelated, and $\mathbf{R_s} = \mathbb{E}\left[\mathbf{ss}^H\right]$ and $\mathbf{R_n} = \mathbb{E}\begin{bmatrix} \mathbf{nn}^H \end{bmatrix}$ are the covariance matrices of the source and noise signals, respectively.  
When the sound sources and noise components are independent and identically distributed (i.i.d.), these matrices reduce to $\mathbf{R_s} = \sigma_s^2 \mathbf{I}_Q$ and $\mathbf{R_n} = \sigma_n^2 \mathbf{I}_M$, where $\sigma_s^2 $ and $\sigma_n^2 $ are the source and noise powers, respectively, and $\mathbf{I}_Q$ and $\mathbf{I}_M$ are the identity matrices of sizes $Q$ and $M$, respectively. 
In this case, (\ref{eq:eps_bin_2}) can be further simplified to
\begin{align}
\nonumber
\epsilon^{l,r} = & \sigma_s^2 \norm{ [\mathbf{c}^{l,r}]^H \mathbf{V}  - [\mathbf{h}^{l,r}]^T }_2^2 +  \sigma_n^2 \norm{ [\mathbf{c}^{l,r}]^H }_2^2 \\
= & \sigma_s^2 \norm{ \mathbf{V}^H \mathbf{c}^{l,r} - [\mathbf{h}^{l,r}]^* }_2^2 +  \sigma_n^2 \norm{ \mathbf{c}^{l,r} }_2^2,
\label{eq:eps_bin_3}
\end{align}
where $\norm{\cdot}_2$ is the $l^2$-norm, and the second term in (\ref{eq:eps_bin_3}) can be interpreted as Tikhonov-regularization \cite{tikhonov2013numerical}. 
The following derivations of BSM utilize the simplified model in (\ref{eq:eps_bin_3}), which requires less information on the sound field compared to the model in (\ref{eq:eps_bin_2}). While this simplified model may not always hold in practice, it will be shown later in Section~\ref{subsec:BSM_valid_arb_cov} that using (\ref{eq:eps_bin_3}) may not limit the practical use of BSM, and that the formulations do hold for more general sound fields in (\ref{eq:eps_bin_2}).

Next, (\ref{eq:eps_bin_3}) is minimized over the filters $\mathbf{c}^{l,r}$ for each ear separately, in order to produce an accurate binaural signal in the MSE sense:
\begin{align}
	\mathbf{c}^{l,r}_{\text{BSM}} = \text{arg}\,\underset{\mathbf{c}^{l,r}}{\text{min}} \,\epsilon^{l,r},
	\label{eq:c_BSM_argmin}
\end{align}
where $\mathbf{c}^{l,r}_{\text{BSM}}$ are the optimal filters in the MSE sense for the left and right ears, given by \cite{tikhonov2013numerical}:
\begin{align}
\mathbf{c}^{l,r}_{\text{BSM}} = (\mathbf{VV}^H + \frac{\sigma_n^2}{\sigma_s^2} \mathbf{I}_M)^{-1} \mathbf{V} [\mathbf{h}^{l,r}]^*.
\label{eq:c_BSM}
\end{align}
Finally, binaural reproduction with the BSM method can be performed by substituting (\ref{eq:c_BSM}) in (\ref{eq:z_lr}):
\begin{align}
\hat{p}_{\text{BSM}}^{l,r} = [\mathbf{c}^{l,r}_{\text{BSM}}]^H \mathbf{x}. 
\label{eq:p_BSM}
\end{align}

Note that no specific constraints were imposed on the array geometry, and thus, the reproduction in (\ref{eq:p_BSM}) is suitable for any array design. However, the performance of BSM greatly depends on the specific array configuration that is being used. Section~\ref{sec:simulations} proposes several evaluation metrics for the accuracy of BSM with a specific array configuration. 
Further note that the calculation of the BSM filters in (\ref{eq:c_BSM}) requires specific sound-field parameters: the signal and noise powers, the number of assumed sources and their DOAs. If these parameters are known or estimated, they can be used to design a signal-dependent BSM solution. However, in this work, a signal-independent approach is studied and a method for generalizing the BSM solution to arbitrary sound fields is described in the following section, including clear guidelines for how to set these parameters.


\subsection{BFBR as a Special Case of BSM}
In this subsection, the BSM solution in (\ref{eq:c_BSM}) will be interpreted using beamformer analysis, which will produce a design for BFBR according to the BSM approach. 
Note that the filters $\mathbf{c}^{l,r}_{\text{BSM}}$ in (\ref{eq:c_BSM}) can be rewritten as
\begin{align}
	\mathbf{c}^{l,r}_{\text{BSM}} = \mathbf{W} [\mathbf{h}^{l,r}]^* ,
	\label{eq:c_BSM_as_BFBR}
\end{align}
where $\mathbf{W} = (\mathbf{VV}^H + \frac{\sigma_n^2}{\sigma_s^2} \mathbf{I}_M)^{-1} \mathbf{V}$ is an $M\times Q$ matrix with columns $\{\mathbf{w}_q\}_{q=1}^{Q}$ that can be interpreted as beamformers designed according to each of the $Q$ signals. 
Substituting (\ref{eq:c_BSM_as_BFBR}) in (\ref{eq:p_BSM}) produces:
\begin{align}
	\nonumber
	\hat{p}_{\text{BSM}}^{l,r} & = [\mathbf{h}^{l,r}]^T \mathbf{W}^H \mathbf{x} \\
	& = \sum_{q=1}^{Q} y_{\text{BSM}}(\theta_q, \phi_q) h^{l,r}(\theta_q, \phi_q),
	\label{eq:p_BSM_as_BFBR}
\end{align}
where 
\begin{align}
	y_{\text{BSM}}(\theta_q, \phi_q) = \mathbf{w}_q^H \mathbf{x}, \quad \forall q=1,\ldots,Q.
	\label{eq:y_BSM}
\end{align}
Notice the similarity between (\ref{eq:y_BSM}) and (\ref{eq:beamforming}), which shows how to design a BFBR system that coincides with BSM reproduction, as outlined below.
\begin{enumerate}
	\item The number of beamformers $D$ should be equal to the number of assumed sources in BSM design, i.e., $D=Q$.
	\item The $d$-th beamformer weights $\mathbf{w}_d$ in (\ref{eq:w_d}) should be equal to the $d$-th column of $\mathbf{W}$ in (\ref{eq:c_BSM_as_BFBR}) for $d=1,\ldots,Q$. This defines $y(\theta_d, \phi_d)$ in (\ref{eq:BFBR}) according to (\ref{eq:y_BSM}).
	\item The scaling factors $\{\alpha_d\}_{d=1}^{Q}$ in (\ref{eq:BFBR}) are all set to one, such that (\ref{eq:BFBR}) and (\ref{eq:p_BSM_as_BFBR}) match.
\end{enumerate}

The structure of $\mathbf{W}$ corresponds to the MMSE processor, described, for instance, in \cite{van2004optimum}, i.e., it produces source signal estimates with minimum MSE. These estimates are then multiplied by the HRTF of the corresponding direction, as can be seen in (\ref{eq:c_BSM_as_BFBR}). 
Furthermore, since the sources are assumed to be spatially white, this MMSE processor is a scaled version of the MVDR beamformer that produces MMSE estimates instead of distortionless responses \cite{van2004optimum}. 
In the following sections, the quality of BSM reproduction will be studied objectively and subjectively, in a way that enables its use with a BFBR design.

\section{Generalization of BSM Design to Arbitrary Sound Fields}
\label{sec:theoretical_analysis}
The BSM method presented above was designed to reproduce binaural signals for a sound field composed of a finite set of sources with known DOAs. 
In many applications, however, these assumptions may not hold or source information may not be available, thus limiting the use of the method.  
In addition, it was assumed that the sound sources are uncorrelated, comprising another limitation of applying BSM in real-life acoustic environments which include reverberation, and where the assumption of uncorrelated sources is usually violated. 
In light of these limitations, this section provides conditions under which BSM may accurately reproduce binaural signals even for complex acoustic environments with sufficiently high signal-to-noise (SNR) ratio. 
No assumptions on specific array geometry will be used, and thus these conditions can be useful for any arbitrary array. First, the validity of BSM will be shown for sound fields composed of an arbitrary set of sound sources, i.e, an arbitrary number of sources with arbitrary DOAs. Then, the assumption of i.i.d. sound sources will be relaxed where the validity of BSM will be shown for arbitrary covariance of source signals. 

\subsection{BSM Design Valid for Arbitrary Set of Sources}
\label{subsec:BSMcondition}
To analyze the BSM method for sound fields composed of an arbitrary set of sound sources, the BSM reproduction error in (\ref{eq:eps_bin_3}) is studied. More specifically, since the right term in (\ref{eq:eps_bin_3}) is a regularization term, the following analysis includes only the left term, that describes the binaural signal error excluding the noise, and it is thus assumed that the SNR is sufficiently high, i.e. $\sigma_s^2 \gg \sigma_n^2$. Next, assume that the BSM filters $\mathbf{c}_{\text{BSM}}^{l,r}$ produce a sufficiently small error such that it is effectively zero:
\begin{align}
	\norm{ \mathbf{V}^H \mathbf{c}_{\text{BSM}}^{l,r} - [\mathbf{h}^{l,r}]^* }_2^2 = 0.
	\label{eq:HRTF_matching_1}
\end{align}
Since the $l^2$-norm in (\ref{eq:HRTF_matching_1}) is zero, the following holds:
\begin{align}
	\mathbf{V}^H \mathbf{c}_{\text{BSM}}^{l,r} = [\mathbf{h}^{l,r}]^* .
	\label{eq:HRTF_matching_2}
\end{align}

Equation (\ref{eq:HRTF_matching_2}) depends on the specific source directions of the assumed sound-field. In order to eliminate this dependency and thus generalize the validity of BSM, the analysis of  (\ref{eq:HRTF_matching_2}) will be performed using SH formulation, which facilitates this goal, as shown next. For this purpose, further assume that the HRTFs, $h^{l,r}(\theta, \phi)$, and the array transfer functions (ATFs), $\{ v(\mathbf{d}_m; \theta, \phi) \}_{m=1}^{M}$, are order limited in the SH domain \cite{rafaely2015fundamentals} up to orders of $N_H$ and $N_V$, respectively. 
In this case, (\ref{eq:HRTF_matching_2}) can be described using the SH basis functions as \cite{rafaely2015fundamentals}:
\begin{align}
	\mathbf{Y}_{N_{V}} \mathbf{V}^H_{\mathbf{nm}} \mathbf{c}_{\text{BSM}}^{l,r} = \mathbf{Y}_{N_{H}} [\mathbf{h}^{l,r}_{\mathbf{nm}}]^*,
	\label{eq:HRTF_matching_2_SH_domain}
\end{align}
where 
\begin{align}
	\resizebox{1\hsize}{!}{%
		$\mathbf{Y}_{ N } = \begin{bmatrix}
			Y_0^0(\theta_1, \phi_1) & Y_1^{-1}(\theta_1, \phi_1) & Y_1^{0}(\theta_1, \phi_1) & \ldots & Y_{ N }^{ N }(\theta_1, \phi_1) \\
			Y_0^0(\theta_2, \phi_2) & Y_1^{-1}(\theta_2, \phi_2) & Y_1^{0}(\theta_2, \phi_2) & \ldots & Y_{ N }^{ N }(\theta_2, \phi_2) \\
			\vdots & \vdots & \vdots & \ddots & \vdots \\
			Y_0^0(\theta_Q, \phi_Q) & Y_1^{-1}(\theta_Q, \phi_Q) & Y_1^{0}(\theta_Q, \phi_Q) & \ldots & Y_{ N }^{ N }(\theta_Q, \phi_Q)
		\end{bmatrix}$
	}
	\label{eq:Y_V_matrix}
\end{align}
is a $Q\times (N +1)^2$ matrix holding the SH functions $Y_n^m(\theta, \phi)$ of order $n$ and degree $m$ \cite{rafaely2015fundamentals}, 
$\mathbf{V}^H_{\mathbf{nm}}$ is an $( N_V +1)^2\times M$ matrix with the $m$-th column holding the SFT coefficients of the complex conjugate of the transfer function $ v(\mathbf{d}_m; \theta, \phi)$ up to order $N_V$ for $m=1,\ldots,M$, and $\mathbf{h}^{l,r}_{\mathbf{nm}}$ is a vector of size $( N_H +1)^2$ holding the SFT coefficients of the HRTFs up to order $N_H$. 

Next, in order to generalize (\ref{eq:HRTF_matching_2_SH_domain}) to an arbitrary set of sources, we aim to omit the dependency on the set of sources in the design $\{(\theta_q, \phi_q)\}_{q=1}^{Q}$, by omitting matrices $\mathbf{Y}_{N_{V}}$ and $\mathbf{Y}_{N_{H}}$ from (\ref{eq:HRTF_matching_2_SH_domain}). 
Furthermore, this generalization should support the matching of ATFs to the full SH representation of the HRTFs, and hence it is assumed that $N_V \ge N_H $ such that (\ref{eq:HRTF_matching_2_SH_domain}) can be rewritten as
\begin{align}
	\mathbf{Y}_{N_{V}} \mathbf{V}^H_{\mathbf{nm}} \mathbf{c}_{\text{BSM}}^{l,r} = \mathbf{Y}_{N_{V}} \begin{bmatrix}
		[\mathbf{h}^{l,r}_{\mathbf{nm}}]^* \\ \mathbf{0}
	\end{bmatrix},
	\label{eq:HRTF_matching_2_SH_domain_zp}
\end{align}
where $\begin{bmatrix}
	[\mathbf{h}^{l,r}_{\mathbf{nm}}]^* \\ \mathbf{0}
\end{bmatrix}$ is a zero-padded version of $[\mathbf{h}^{l,r}_{\mathbf{nm}}]^*$ to a total length of $\big( N_V + 1 \big)^2$. 
Then, omitting matrix $\mathbf{Y}_{N_{V}}$ from (\ref{eq:HRTF_matching_2_SH_domain_zp}) can be performed assuming that the following holds: 
\begin{align}
	[\mathbf{Y}_{N_{V}}]^{\dagger} \mathbf{Y}_{N_{V}} = \mathbf{I}_{(N_V + 1)^2},
	\label{eq:Y_dagger}
\end{align}
where $[\mathbf{Y}_{N_{V}}]^{\dagger} $ is the pseudo-inverse of $\mathbf{Y}_{N_{V}}$. Multiplying (\ref{eq:HRTF_matching_2_SH_domain_zp}) from the left by $[\mathbf{Y}_{N_{V}}]^{\dagger}$ and substituting (\ref{eq:Y_dagger}) results in
\begin{align}
	\mathbf{V}^H_{\mathbf{nm}} \mathbf{c}_{\text{BSM}}^{l,r} =  \begin{bmatrix}
		[\mathbf{h}^{l,r}_{\mathbf{nm}}]^* \\ \mathbf{0}
	\end{bmatrix}.
	\label{eq:HRTF_matching_2_SH_domain_2}
\end{align}
In order to guarantee that (\ref{eq:Y_dagger}) is satisfied, the number of sources in the design should satisfy  $Q \ge \big( N_V + 1 \big)^2$ and the DOA set $\{(\theta_q, \phi_q)\}_{q=1}^{Q}$ which comprise $\mathbf{Y}_{N_{V}}$ should be determined according to a sampling scheme on the sphere that is aliasing-free up to an SH order of $N_V$ \cite{rafaely2015fundamentals}. 
While this may seem like a hard condition, Section~\ref{sec:simulations} shows that it can be readily satisfied with the studied array (see Fig.~\ref{fig:effectiveSH_order} for example). 
Finally, (\ref{eq:HRTF_matching_2_SH_domain_2}) can now be multiplied from the left by an SH matrix similar to (\ref{eq:Y_V_matrix}), but with any arbitrary set of angles, $\{(\theta_a, \phi_a)\}_{a=1}^{A}$, denoted $\mathbf{Y}'_{N_{V}}$, leading to:
\begin{align}
	\mathbf{Y}'_{N_{V}} \mathbf{V}^H_{\mathbf{nm}} \mathbf{c}_{\text{BSM}}^{l,r} = \mathbf{Y}'_{N_{V}} \begin{bmatrix}
		[\mathbf{h}^{l,r}_{\mathbf{nm}}]^* \\ \mathbf{0}
	\end{bmatrix}.
	\label{eq:HRTF_matching_2_SH_domain_zp_2}
\end{align}
The equality in (\ref{eq:HRTF_matching_2_SH_domain_zp_2}) means that the BSM filters that were designed for sources corresponding to $\{(\theta_q, \phi_q)\}_{q=1}^{Q}$ can be used to match binaural signals corresponding to any arbitrary set. 
Finally, note that the SH formulation was used here to facilitate the theoretical analysis, but in this work, the calculation of the BSM filters will be based on space-domain formulation, as described in (\ref{eq:c_BSM}).

\subsection{BSM Design Valid for Arbitrary Covariance of Source Signals}
\label{subsec:BSM_valid_arb_cov}
Next, the validity of BSM filters designed according to (\ref{eq:c_BSM}) is analyzed for source signals with an arbitrary covariance matrix $\mathbf{R_s}$ in (\ref{eq:eps_bin_2}). 
Once again, excluding the noise from the analysis reduces (\ref{eq:eps_bin_2}) to:
\begin{align}
    \resizebox{1\hsize}{!}{%
        $
    	\epsilon^{l,r} \approx  \big([ \mathbf{c}_{\text{BSM}}^{l,r} ]^H \mathbf{V} - [\mathbf{h}^{l,r}]^T\big) \mathbf{R_s} \big([ \mathbf{c}_{\text{BSM}}^{l,r} ]^H \mathbf{V} - [\mathbf{h}^{l,r}]^T\big)^H.
        $
     }
	\label{eq:eps_bin_2_highSNR}
\end{align}
Based on the assumptions and derivations in the previous section, substituting (\ref{eq:HRTF_matching_2}), which yields $[\mathbf{h}^{l,r}]^T=[\mathbf{c}_{\text{BSM}}^{l,r} ]^H \mathbf{V}$, in (\ref{eq:eps_bin_2_highSNR}) leads to the binaural signals' MSE, $\epsilon^{l,r}$, being zero, thus extending the validity of BSM to arbitrary $\mathbf{R_s}$. 
However, in the case where (\ref{eq:HRTF_matching_2}) does not hold, such generalization may not be valid. In this case, incorporating $\mathbf{R_s}$ (or its estimate) in the BSM design may reduce the error, but this is out of the scope of this paper and is suggested for future work. Nevertheless, Section~\ref{sec:magLS} will present a perceptually-motivated improvement to BSM at high frequencies, where satisfying (\ref{eq:HRTF_matching_2}) is more challenging.

\subsection{Summary of Conditions for BSM Design Generalization} 
\label{subsec:conditions}
To summarize, the conditions that generalize specific BSM design to hold in more complex sound fields, with an arbitrary source set and with source signals that are not necessarily spatially white, are outlined below. 
\begin{enumerate}
	\item A sufficiently high SNR. 
	\item The ATFs and HRTFs are order limited in the SH domain to orders $N_V $ and $N_H $, respectively.
	\item The number of sources in the design satisfies $Q \ge \big(N_V + 1 \big)^2 $ \cite{rafaely2015fundamentals}. 
	\item The source DOAs in the design, $\{(\theta_q, \phi_q)\}_{q=1}^{Q}$, are determined according to a sampling scheme on the sphere that is aliasing-free up to an SH order of $N_V$ \cite{rafaely2015fundamentals}. 
	\item The maximal SH order of the ATFs is at least as large as that of the HRTFs, i.e., $N_V \ge N_H$.
\end{enumerate}
Since the ATFs and HRTFs are assumed to be known, the BSM design can be performed according to the guidelines above independently of the actual sound field.

\section{Performance Limitations at High Frequencies and a Perceptually-motivated Extension}
\label{sec:magLS}
The previous section theoretically analyzed the BSM method based on the assumption that (\ref{eq:HRTF_matching_2}) is satisfied. 
In this section, it is argued that practical arrays may not satisfy (\ref{eq:HRTF_matching_2}) at high frequencies due to limited spatial resolution, which suggests that the accuracy of BSM may degrade. 
Following that, a possible solution to improve the accuracy of a modified BSM problem at high frequencies will be proposed.  

Recall that the BSM filters are the solution to the minimization in (\ref{eq:c_BSM_argmin}), rewritten here:
\begin{align}
    \resizebox{1\hsize}{!}{%
        $
    	\mathbf{c}^{l,r}_{\text{BSM}} = \text{arg}\,\underset{\mathbf{c}^{l,r}}{\text{min}} \,\big\{\sigma_s^2 \norm{ \mathbf{V}^H \mathbf{c}^{l,r} - [\mathbf{h}^{l,r}]^* }_2^2 +  \sigma_n^2 \norm{ \mathbf{c}^{l,r} }_2^2\big\}.
    	\label{eq:c_BSM_argmin_extended}
        $
    }
\end{align}
The first term in the minimization in (\ref{eq:c_BSM_argmin_extended}) can be interpreted as the error of matching ATFs to the HRTFs, and is directly related to the accuracy of binaural reproduction with BSM. This error will be small for filters $\mathbf{c}^{l,r}$ that satisfy:
\begin{align}
	 \mathbf{V}^H \mathbf{c}^{l,r} \approx [\mathbf{h}^{l,r}]^*,
	 \label{eq:HRTF_matching_approx}
\end{align}
which is a linear system with $Q$ constrains and $M$ degrees of freedom. 

In addition, the previous section showed that in order for the BSM solution to be valid for complex sound fields it is required that the number of sources in the design will satisfy: 
\begin{align}
	Q \ge (N_H+1)^2
	\label{eq:cond3}
\end{align}
(see condition 3). Since the effective SH order of the HRTFs, $N_H$, generally increases with frequency \cite{zhang2010insights}, it may be desired to choose a sufficiently large $Q$ for (\ref{eq:cond3}) to hold at high frequencies. 
However, this may cause $Q$ to be much larger than the number of microphones $M$ in practical arrays, especially in the high frequency range \cite{zhang2010insights}. In this case, (\ref{eq:HRTF_matching_approx}) will be an overdetermined system, such that the solution in (\ref{eq:c_BSM_argmin_extended}) may produce a relatively large error, and thus, the accuracy of BSM may significantly degrade.

With the aim of reducing the effect of this loss of accuracy at high frequencies, a perceptually-motivated alternative is proposed for BSM in this work. It is based on the precept that inter-aural level differences (ILD) are more important than inter-aural time differences (ITD) for spatial perception at high frequencies \cite{ITD_1500_1,ITD_1500_2,ITD_1500_3,ITD_less_importnt_high_freq}. Hence, at high frequencies, the MSE in (\ref{eq:eps_bin_3}) is relaxed by replacing the complex binaural signal matching with the matching of absolute values:
\begin{align}
	\epsilon_{\text{abs}}^{l,r} = \sigma_s^2 \norm{ |\mathbf{V}^H \mathbf{c}^{l,r}| - |[\mathbf{h}^{l,r}]^*| }_2^2 +  \sigma_n^2 \norm{\mathbf{c}^{l,r} }_2^2.
	\label{eq:eps_bin_magLS}
\end{align}
Thus, with (\ref{eq:eps_bin_magLS}), the aim is to match only the magnitude values of the HRTFs, ignoring the phase. This coincides with the understanding that for spatial perception, preserving the magnitude values of the binaural signal could be useful to improve reproduction of ILD. 
Then, the objective of BSM at these frequencies is to find $\mathbf{c}^{l,r}$ that minimizes (\ref{eq:eps_bin_magLS}), formally written as:
\begin{align}
	\mathbf{c}_{\text{BSM-MagLS}}^{l,r} = \text{arg}\,\underset{{\mathbf{c}}^{l,r}}{\text{min}}\,\epsilon_{\text{abs}}^{l,r}.
	\label{eq:c_MLS_BSM}
\end{align}
The solution to this problem is sometimes referred to as MagLS \cite{MagLS_kassakian2006convex,MagLS_setsompop2008magnitude,MagLS_schorkhuber2018binaural}. 
Notice that the difference between (\ref{eq:eps_bin_magLS}) and (\ref{eq:eps_bin_3}) is the absolute values on the terms $\mathbf{V}^H \mathbf{c}^{l,r}$ and $[\mathbf{h}^{l,r}]^*$, which reduces the complexity of the problem in (\ref{eq:c_MLS_BSM}), and therefore could lead to reduced errors. 
This new objective is non-convex, and thus, a global minimum may be unachievable. Nonetheless, various approaches for finding a local minimum to (\ref{eq:eps_bin_magLS}) are given in \cite{MagLS_kassakian2006convex} alongside a theoretical analysis of their quality. 

The incorporation of MagLS into the BSM method is motivated by the works in \cite{MagLS_schorkhuber2018binaural,zotter2019ambisonics}, which used MagLS for rendering Ambisonics signals. Moreover, the solution proposed here is similar to the eMagLS method proposed in \cite{e2e_magls}. Similar to \cite{e2e_magls}, it is suggested to incorporate the complex least-squares solution to $\epsilon^{l,r}$ in (\ref{eq:c_BSM_argmin}) below a predefined cutoff frequency for which (\ref{eq:HRTF_matching_approx}) holds, with the MagLS solution to $\epsilon_{\text{abs}}^{l,r}$ in (\ref{eq:c_MLS_BSM}) employed above the cutoff. 
The value of the cutoff frequency can be determined empirically by examining the BSM performance of the specific array being used, and typically it should be roughly above 1.5 kHz where the ILD is more important than the ITD for spatial perception \cite{ITD_1500_1,ITD_1500_2,ITD_1500_3,ITD_less_importnt_high_freq}.

\section{Compensating for Head Rotations with BSM Reproduction}
\label{sec:head_rot}
This section addresses head rotations with BSM reproduction. Two types of rotation and their effect on reproduction are introduced, and a method to compensate for these rotations is proposed. 
Throughout this discussion it is assumed that full information on head orientation is known via a head-tracking device. 
For simplicity, this discussion is focused on yaw rotation along the azimuthal plane, however, the principles extend to all rotational degrees of freedom.

The first rotation type corresponds to head rotations of the listener during the playback stage of binaural reproduction with BSM. 
In this scenario, it is desired to reproduce binaural signals corresponding to an acoustic scene that is fixed with respect to the environment of the listener. 
To illustrate this, Fig.~\ref{fig:acoustic_environment} shows a person wearing a head-mounted device with an embedded microphone array. The signals recorded by the array are then processed to generate binaural signals that are then played back to a remote listener wearing headphones. To enhance the immersion of the listener, his/her head rotations, denoted by red arrows in Fig.~\ref{fig:acoustic_environment}, should be compensated for \cite{begault2001direct}. With BSM reproduction, this can be performed by modifying the HRTF vector $\mathbf{h}^{l,r}$ in (\ref{eq:c_BSM}) to embody the correct rotated HRTFs. These are denoted by $\mathbf{h}_{\text{rot}}^{l,r}$ and are given by the following expression:
\begin{align}
	\resizebox{1\hsize}{!}{%
		$\mathbf{h}_{\text{rot}}^{l,r} =\begin{bmatrix}
			h^{l,r}(\theta_1 + \Delta\theta , \phi_1 + \Delta\phi), \ldots, h^{l,r}(\theta_Q + \Delta\theta, \phi_Q + \Delta\phi)
		\end{bmatrix}^T, $
	}
	\label{eq:HRTF_rot}
\end{align}
where $ \Delta\theta$ and $ \Delta\phi$ are the amount of head rotation in degrees, in elevation and azimuth, respectively.

The second rotation type is relevant for a head-mounted array recording an acoustic scene. 
In this scenario, it is desired to reproduce binaural signals that represent an acoustic scene that is fixed with respect to a reference coordinate system within the recording environment. 
However, when the person wearing the recording device rotates his/her head, the recorded acoustic scene rotates in the opposite direction relative to this reference coordinate system. This is illustrated in Fig.~\ref{fig:acoustic_environment} by the gray arrows above the recording person on the left. 
In order to compensate for this rotation, the steering vectors in (\ref{eq:ATF}), which comprise the columns of $\mathbf{V}$ in (\ref{eq:c_BSM}), can be modified as follows:
\begin{align}
	\resizebox{1\hsize}{!}{%
		$ \mathbf{v}^{\text{rot}}_q =\begin{bmatrix}
			v(\Delta\mathbf{d}_1 ; \theta_q, \phi_q), v(\Delta\mathbf{d}_2 ; \theta_q, \phi_q), \ldots, v(\Delta\mathbf{d}_M ; \theta_q, \phi_q)
		\end{bmatrix}^T   $, %
	}
	\label{eq:ATF_rot}
\end{align}
for $q=1,2,\ldots,Q$, where $\Delta\mathbf{d}_m$ is the rotated position of the $m$-th microphone in the array with respect to the reference orientation for $m=1,2,\ldots,M$. 

Finally, notice that both rotation types can occur in the same recording and reproduction session. In this case, both can be compensated for separately by modifying the BSM filters in (\ref{eq:c_BSM}) according to (\ref{eq:HRTF_rot}) and (\ref{eq:ATF_rot}). 
Compensating for head rotations with these equations can potentially degrade the reproduction quality of BSM, as will be shown later. 
This is because the modified BSM filters reproduce binaural signals that correspond to ear positions that may be relatively far from the microphone positions \cite{BSM_I3DA}. This depends on the degree of rotation and on the array configuration. 

\section{Simulation Study of BSM with a Semi-circular Array}
\label{sec:simulations}
Having established the theoretical foundation for BSM performance, this section presents a performance analysis using a microphone array mounted on a rigid sphere with a semi-circle geometry as a case study. While the results are specific to this configuration, the presented evaluation methodology is broadly applicable and thus serves as a template for assessing the performance of BSM and BSM-MagLS with other array geometries. This provides a valuable tool for selecting array geometry during the design phase.

\subsection{Experimental Setup}
\label{subsec:simulation_setup}
The array which will be employed throughout this section is comprised of $M=6$ microphones distributed on a semi-circle that is mounted on a rigid sphere. The spherical coordinates of this array are given by $ r_m = 10\,\text{cm} $, $ \theta_m = \frac{\pi}{2} \,$rad, and $ \phi_m = \frac{\pi}{2} - \frac{\pi (m-1)}{M-1} \,$rad for $m=1,\ldots,M$. 
The ATFs for this array were calculated in the SH domain up to an order of $N=30$, as described in Section 4.2 in \cite{rafaely2015fundamentals} for rigid spheres. 
In addition, the HRTFs studied here are from the measured Neumann KU100 manikin from the Cologne database \cite{CologneDatabase} with a sampling frequency of $48\,$kHz and a Lebdev sampling scheme consisting of 2702 points. 

\subsection{BSM Accuracy under Static Conditions}
\label{subsec:simulation_static}
In this subsection, the full least-squares solution of BSM is analyzed, i.e., without its MagLS extension. For this purpose, the following normalized error measure was defined:
\begin{align}
	\bar{\epsilon}^{\,l,r}(k) = \frac{ \mathbb{E}  [ | p^{l,r}(k) - \hat{p}^{l,r}(k) | ]^2  } { \mathbb{E} [ | p^{l,r}(k) |^2 ]  }.
	\label{eq:eps_norm_1}
\end{align}
Substituting (\ref{eq:binau_repr_space_L_PWs}) and (\ref{eq:p_BSM}) in (\ref{eq:eps_norm_1}) (using (\ref{eq:array_meas_VanTrees}) and (\ref{eq:c_BSM})) results in the following more explicit expression:
\begin{align}
    \resizebox{1\hsize}{!}{%
        $
    	\bar{\epsilon}^{\,l,r}(k) = \frac{ \sigma_s^2 \norm{ \mathbf{V}^H \mathbf{c}_{\text{BSM}}^{l,r} - [\mathbf{h}^{l,r}]^* }_2^2 +  \sigma_n^2 \norm{ \mathbf{c}_{\text{BSM}}^{l,r} }_2^2 } { \sigma_s^2 \norm{  [\mathbf{h}^{l,r}]^* }_2^2  }.
        $
    }
	\label{eq:eps_norm_2}
\end{align}
This is the analytical error of BSM reproduction at each ear, when the acoustic environment is comprised of $Q$ uncorrelated sources and white noise, and for the array measurement model in (\ref{eq:array_meas_VanTrees}).

\begin{figure}
	\centering
	\begin{minipage}{1\columnwidth}	
		\centering
		\begin{tikzpicture}
			\node (img)  {\includegraphics[width=0.99\columnwidth,trim={0cm 0cm 0cm 0cm},clip]{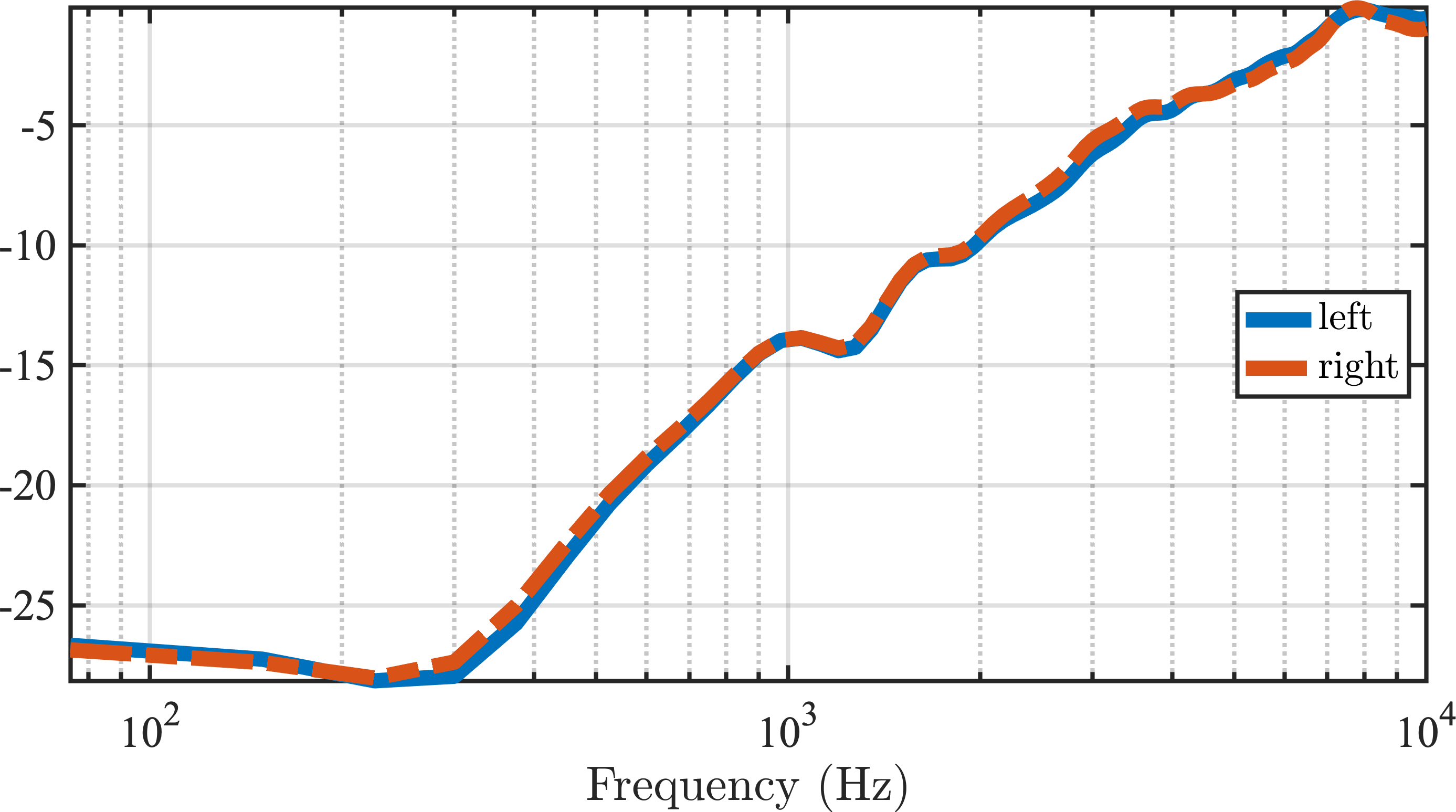}};
			\node[overlay, left=of img, node distance=0cm, rotate=90, anchor=center,yshift=-0.9cm,font=\color{black}] {\small \small $\bar{\epsilon}^{\,l,r}(k)$  [dB]};
		\end{tikzpicture}	
	\end{minipage}	
	\begin{minipage}{1\columnwidth}
		\centering
		\caption{The analytical error of BSM reproduction in (\ref{eq:eps_norm_2}) for both ears calculated with the semi-circular array with $M=6$ microphones and for the BSM design parameters described in subsections~\ref{subsec:simulation_setup} and \ref{subsec:simulation_static} \label{fig:eps_norm_lr_static}	}		
	\end{minipage}
	
\end{figure}

The error in (\ref{eq:eps_norm_2}) was calculated for $Q=240$ source directions, corresponding to a nearly-uniform spiral scheme \cite{saff1997distributing}, and with a $20\,$dB SNR, by setting $\sigma_s^2$ and $\sigma_n^2$ accordingly. The error is presented for frequencies in the range of $[75, 10000]\,$Hz with $75\,$Hz resolution in Fig.~\ref{fig:eps_norm_lr_static}. 
Notice that the error is relatively low, i.e., below 10\,dB, for frequencies below approximately 1.5\,kHz for both ears. Hence, in this frequency range the BSM method is expected to reproduce the acoustic scene accurately. However, the reproduction error increases for higher frequencies, and above approximately 2\,kHz it has become very large, such that the reproduction is expected to be poor. 
As was explained in Section~\ref{sec:magLS}, BSM accuracy may degrade at high frequencies due to the increase in the maximal SH orders of the HRTFs, and hence this is studied next.

\subsection{Effective SH Order of the ATFs and the HRTFs}
\label{subsec:effSH_order}
The purpose of this subsection is to analyze the effective SH orders of the ATFs ($N_V$) and HRTFs ($N_H$), which are part of the conditions for BSM generalization that are described in Section~\ref{subsec:conditions}. 
In order to study the effective SH order of a function on the sphere $f(\theta,\phi)$, the following cumulative energy measure up to the $N$-th SH order is defined as:
\begin{align}
	E(N) = \sum_{n=0}^{N}\sum_{m=-n}^{n} | f_{nm} |^2, 
	\label{eq:E_N}
\end{align}
where $f_{nm}$ is the SFT coefficients of $f(\theta,\phi)$ of order $n$ and degree $m$. Next, (\ref{eq:E_N}) is normalized according to:
\begin{align}
	\bar{E}(N) = \frac{ E(N)  }{\underset{N}{ \text{max}\,}E(N)}.
	\label{eq:E_N_norm}
\end{align}
Finally, the effective SH order of $f(\theta,\phi)$ corresponding to $X\%$ of the energy is defined as \cite{ben2019efficient}:
\begin{align}
	b_{X} = \underset{N}{\text{min}} \big\{ \big| 	\bar{E}(N) - \frac{X}{100} \big| \big\}.
	\label{eq:b_X}
\end{align}
This measure is studied next with the HRTFs and ATFs that were used to calculate the BSM errors presented in Fig.~\ref{fig:eps_norm_lr_static}.

\begin{figure}
	\centering
		\begin{tikzpicture}
			\node (img)  {\includegraphics[width=0.99\columnwidth,trim={0cm 0cm 0cm 0cm},clip]{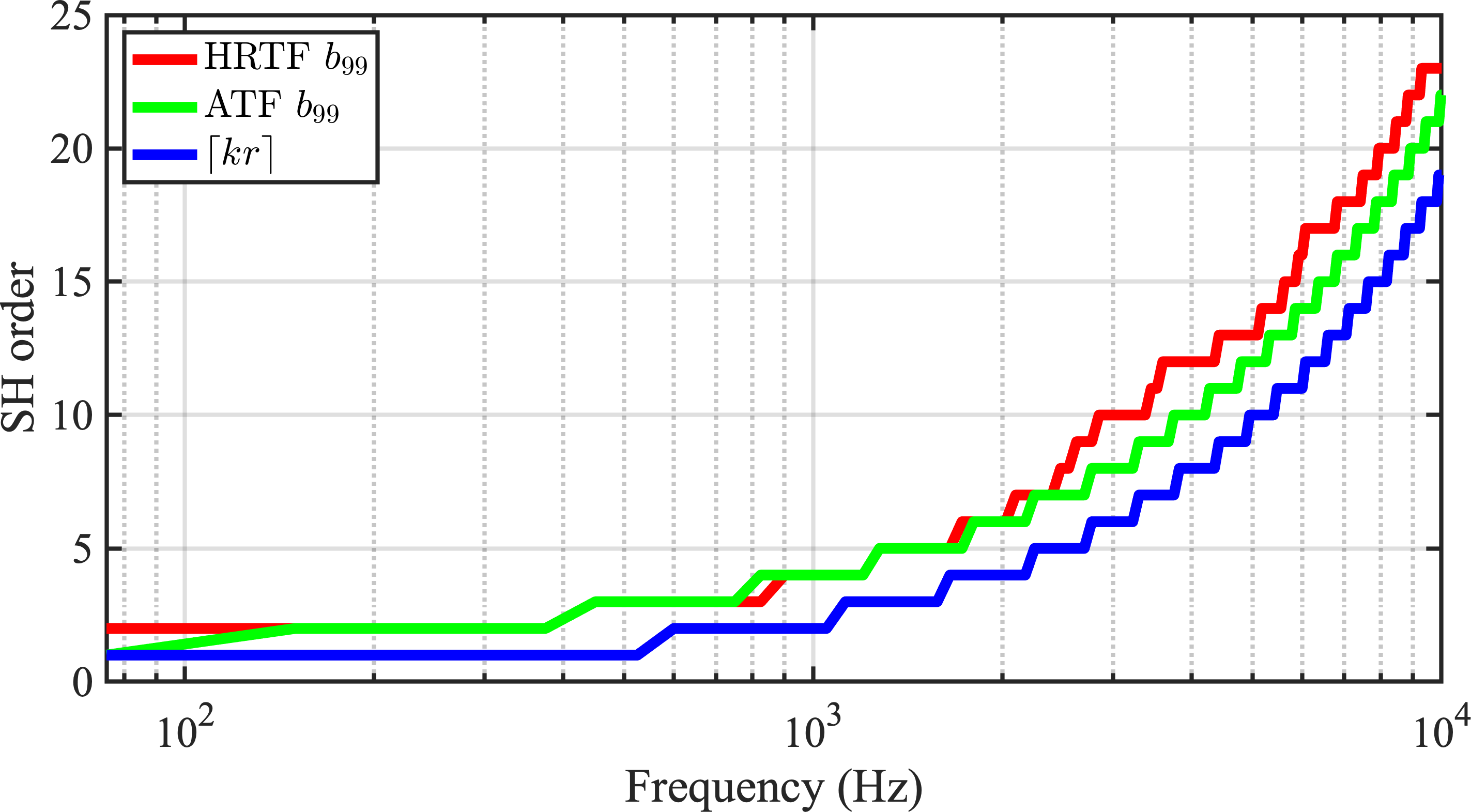}};
		\end{tikzpicture}	
	\caption{Effective SH order according to the measure $b_{99}$ in (\ref{eq:b_X}) of left ear HRTF and of the ATF corresponding to the 1-st microphone of the semi-circular array with $M=6$ microphones. In addition, $\lceil kr \rceil$ is shown for reference, calculated using the array radius.}
	\label{fig:effectiveSH_order}
\end{figure}

For this purpose, the SFT of the left ear HRTF and the ATF corresponding to microphone $m=1$ were calculated with the 2702 Lebdev sampling points. Next, $b_{99}$ was calculated for frequencies in the range of $[75, 10000]\,$Hz with $75\,$Hz resolution, and is presented in Fig.~\ref{fig:effectiveSH_order}. The figure also shows $\lceil kr \rceil$ calculated using the array radius, for reference. 
Notice the rapid increase of the maximal SH order of the HRTF with frequency. This may lead to (\ref{eq:cond3}) not being satisfied at high frequencies, such that (\ref{eq:HRTF_matching_approx}) is an under-determined system, which may explain the relatively large errors in Fig.~\ref{fig:eps_norm_lr_static}.  
In addition, at frequencies where $b_{99}$ of the HRTF is larger than that of the ATF, some HRTF components may not be reconstructed with sufficient accuracy, which may further explain the decrease in accuracy. 
Since the frequency range of accurate BSM reproduction is relatively limited, its MagLS extension, which may potentially increase this range, will be studied next. 

\subsection{MagLS Extension of BSM and Head Rotation Compensation}
\label{subsec:sim_MLS_head_rot}
This part studies the MagLS extension of BSM described in Section~\ref{sec:magLS}, including head rotation compensation. 
In order to calculate the MagLS solution, (\ref{eq:eps_bin_magLS}) was minimized over $\mathbf{c}^{l,r}$, as described in (\ref{eq:c_MLS_BSM}), in accordance with the variable exchange method presented in \cite{MagLS_kassakian2006convex} (Section 5.3.1). 
This iterative method was performed with an initial phase of $\frac{\pi}{2}$, tolerance of $10^{-20}$ and a maximum of $10^5$ iterations. 
In order to focus on magnitude reproduction, the MagLS solution was calculated with a cutoff frequency of 0\,Hz, i.e., for the entire frequency range. 
In addition, the normalized error in (\ref{eq:eps_norm_2}) was modified for this study to capture only errors in magnitude, by incorporating (\ref{eq:eps_bin_magLS})
\begin{align}
    \resizebox{1\hsize}{!}{%
        $
    	\bar{\epsilon}^{\,l,r}_{\text{abs}}(k) = \frac{ \sigma_s^2 \norm{ \big|\mathbf{V}^H \mathbf{c}^{l,r} \big| - \big|[\mathbf{h}^{l,r}]^*\big| }_2^2 +  \sigma_n^2 \norm{ \mathbf{c}^{l,r} }_2^2 } { \sigma_s^2 \norm{  [\mathbf{h}^{l,r}]^* }_2^2  }.
        $
    }
	\label{eq:eps_norm_mag}
\end{align}

\begin{figure}
	\centering
	\begin{minipage}{1\columnwidth}	
		\centering
		\begin{tikzpicture}
			\node (img)  {\includegraphics[width=0.99\columnwidth,trim={0cm 0cm 0cm 0cm},clip]{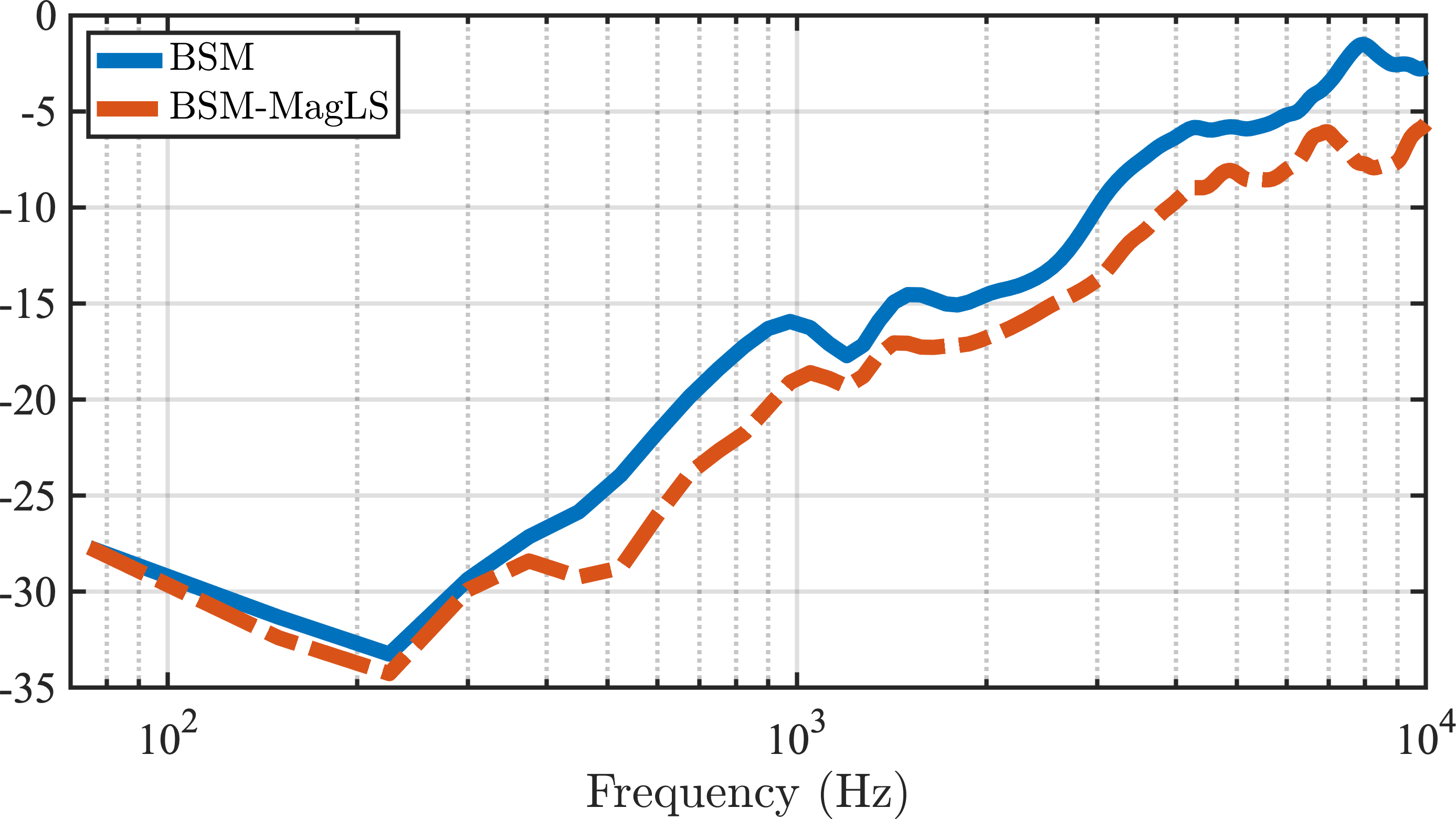}};
			\node[overlay, left=of img, node distance=0cm, rotate=90, anchor=center,yshift=-0.9cm,font=\color{black}] {\small $\bar{\epsilon}^{\,l}_{\text{abs}}(k)$  [dB]};
		\end{tikzpicture}	
	\end{minipage}	
	\begin{minipage}{1\columnwidth}
		\centering
		\caption{The analytical error of magnitude reproduction using BSM and MagLS-BSM reproductions in (\ref{eq:eps_norm_mag}). The error is presented for the left ear signal, and calculated with the semi-circular array with $M=6$ microphones and for the BSM design parameters described in subsection.~\ref{subsec:effSH_order}. \label{fig:eps_abs_norm_l_static}	}		
	\end{minipage}
	
\end{figure}

Figure~\ref{fig:eps_abs_norm_l_static} presents the magnitude errors of the left ear with the original BSM filters $\mathbf{c}_{\text{BSM}}^{l}$ and the BSM-MagLS filters $\mathbf{c}_{\text{BSM-MagLS}}^{l}$. Since the left and rights ear errors are very similar, only the left ear errors are presented here. 
The decrease in error of BSM-MagLS compared to the original BSM averaged across all studied frequencies is $\approx4.2\,$dB and 3.9\,dB for the left and right ears, respectively.
Notice that BSM-MagLS produces smaller errors compared to the original BSM, especially at frequencies in the range [400, 1000]\,Hz and at higher frequencies. This may be attributed to the fact that BSM-MagLS estimates only the magnitude of the HRTFs, while BSM estimates both the magnitude and the phase. However, recall that phase information may be very important for spatial perception, such that the original BSM may produce binaural signals at the lower frequency range which are perceptually better. A more elaborate perceptual study is presented in Section~\ref{sec:list_exp}. 
As a follow-up, these errors are studied next with head rotation compensation.

\begin{figure}
	
	\centering
	\begin{minipage}{1\columnwidth}	
		\centering
		\subfigure[Original]{
			\label{fig:head_array}	
			\begin{tikzpicture}
				\node (img)  {\includegraphics[width=0.33\columnwidth,trim={0.1cm 0.1cm 0cm 0cm},clip]{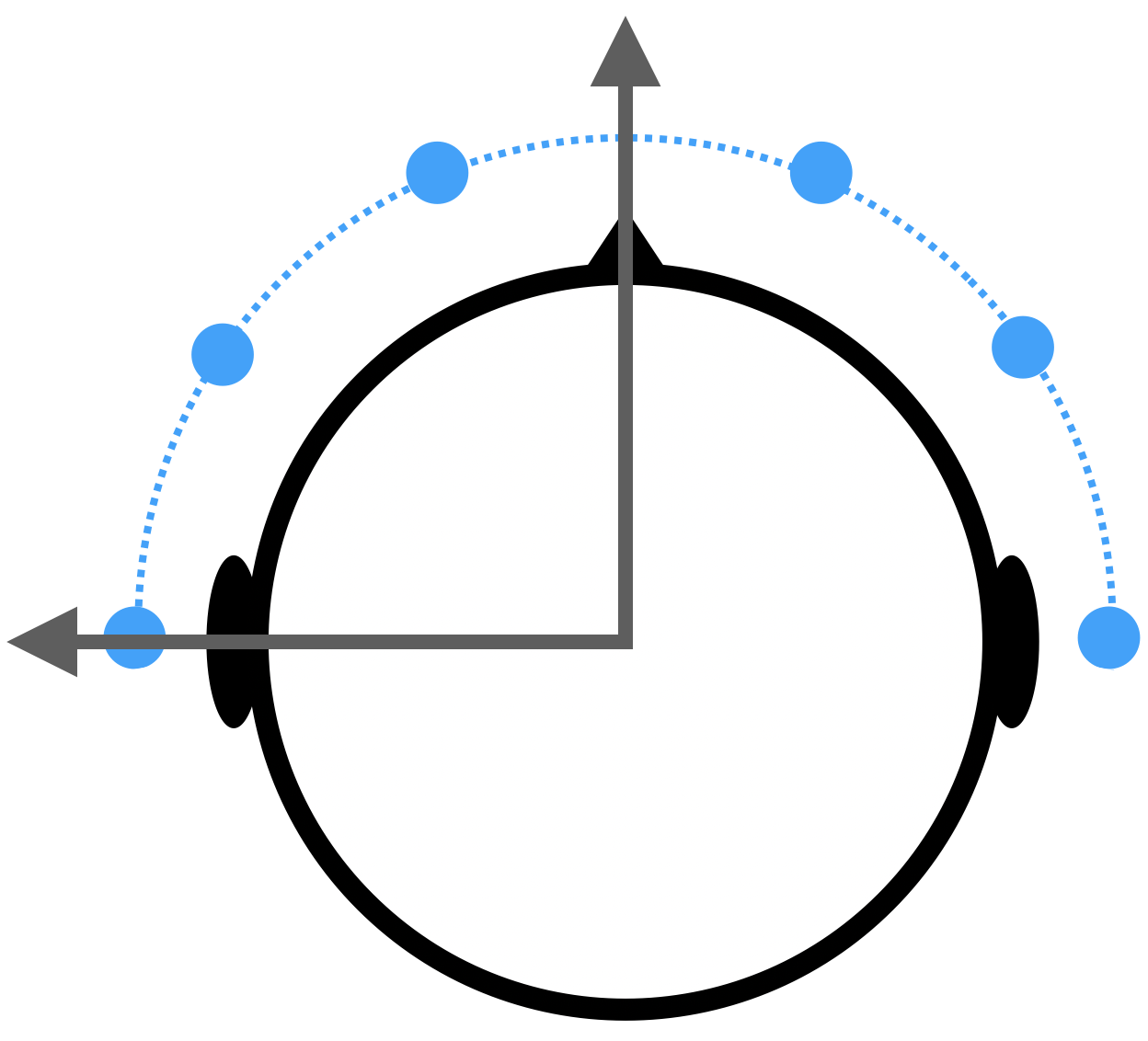}};
				\node[overlay, left=of img, node distance=0cm, rotate=90, anchor=center,yshift=-1.1cm,font=\color{black}] {\small $y$};
				\node[overlay, below=of img, node distance=3cm, yshift=4cm,xshift=0cm,font=\color{black}] {\small $x$};
			\end{tikzpicture}	
		}
		~
		\centering
		\subfigure[Head rotation]{
			\label{fig:head_rotHead}	
			\begin{tikzpicture}
				\node (img)  {\includegraphics[width=0.33\columnwidth,trim={0.1cm 0.1cm 0cm 0cm},clip]{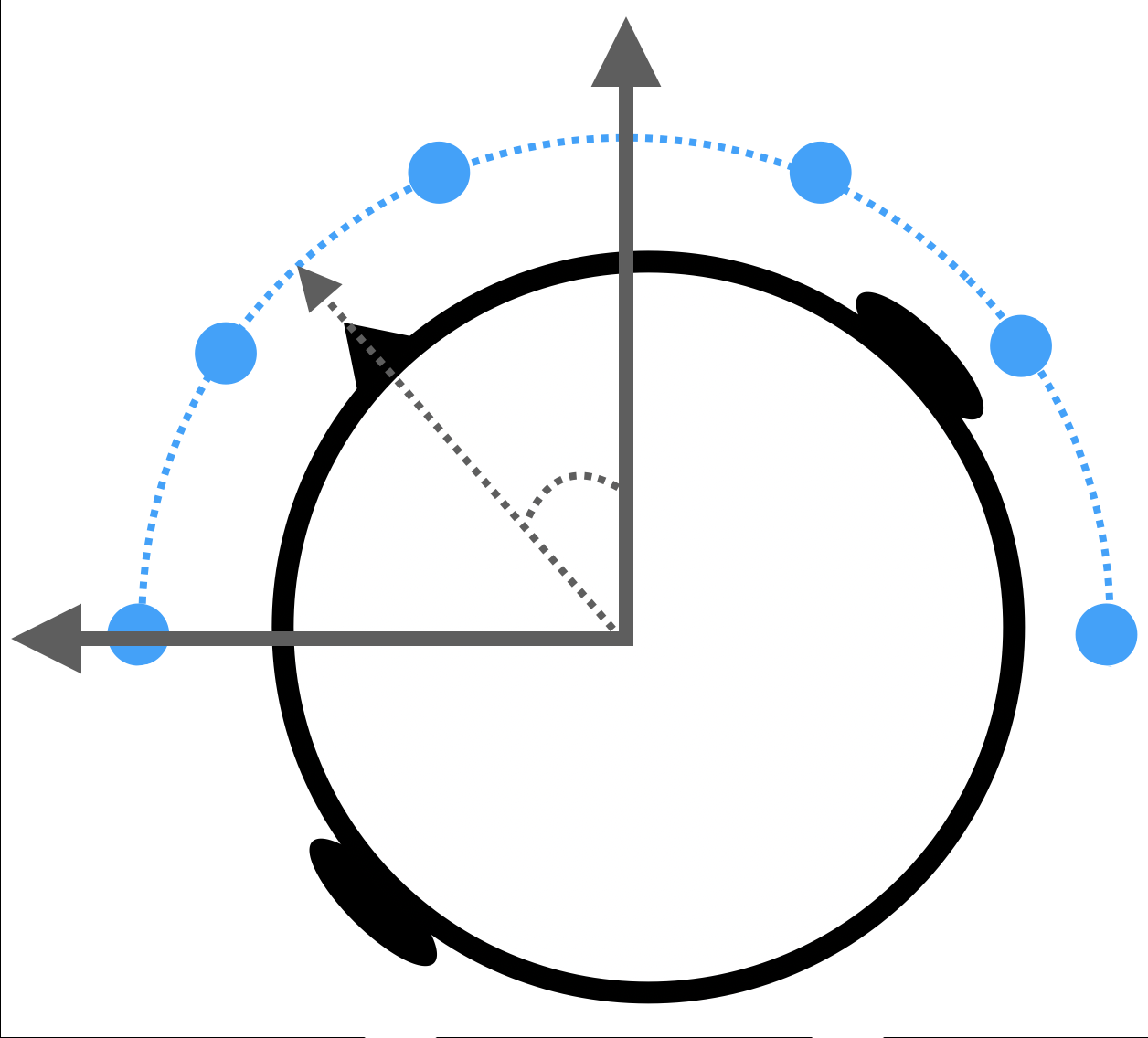}};
				\node[overlay, left=of img, node distance=0cm, rotate=90, anchor=center,yshift=-1.1cm,font=\color{black}] {\small $y$};
				\node[overlay, below=of img, node distance=3cm, yshift=4cm,xshift=0cm,font=\color{black}] {\small $x$};
				\node[overlay, left=of img, node distance=0cm, rotate=0, anchor=center,xshift=2.5cm,yshift=0.2cm,font=\color{black}] {\tiny $\Delta\phi$};
			\end{tikzpicture}	
		}	
	\end{minipage}
	
	\vspace*{-2mm} 
	\begin{minipage}{1\columnwidth}
		\centering
		\caption{Schematic diagram of a head and the semi-circular array with $M=6$ (blue dots). (a) Original orientation, (b) head rotation by $\Delta\phi$ clockwise in azimuth. \label{fig:head_array_diagrams}	}		
	\end{minipage}
	
\end{figure}
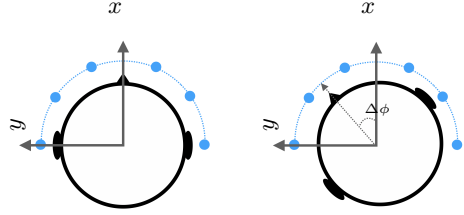

\begin{figure}
	\centering
	\begin{minipage}{1\columnwidth}	
		\centering
		\subfigure[$\Delta\phi$ = 30$^{\circ}$]{
			\label{fig:eps_abs_norm_head_rot_30}	
			\begin{tikzpicture}
				\node (img)  {\includegraphics[width=0.99\columnwidth,trim={0cm 0cm 0cm 0cm},clip]{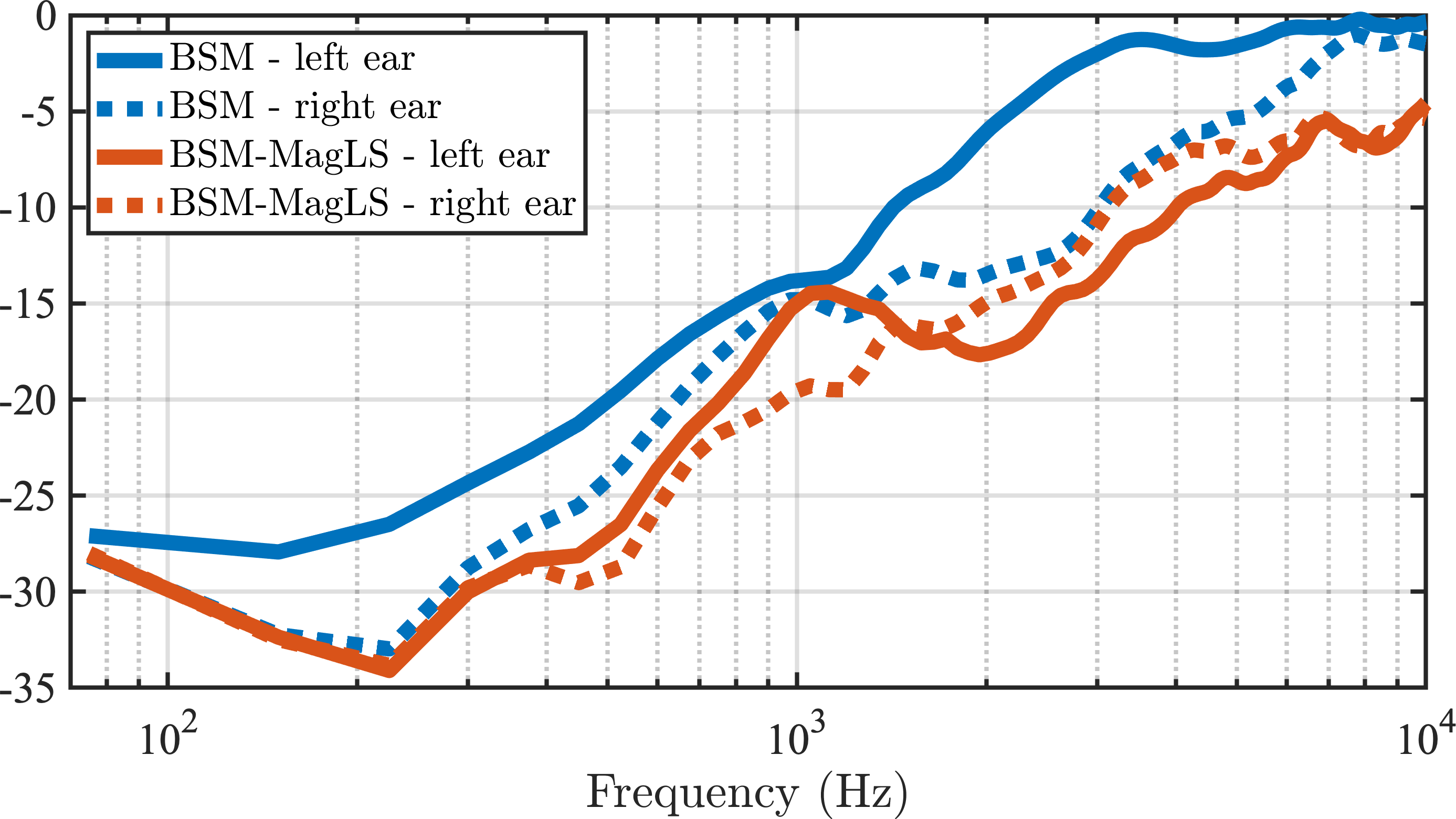}};
				\node[overlay, left=of img, node distance=0cm, rotate=90, anchor=center,yshift=-0.9cm,font=\color{black}] {\small $\bar{\epsilon}^{\,l,r}_{\text{abs}}(k)$  [dB]};
			\end{tikzpicture}	
		}
		~
		\centering
		\subfigure[$\Delta\phi$ = 60$^{\circ}$]{
			\label{fig:eps_abs_norm_head_rot_60}	
			\begin{tikzpicture}
				\node (img)  {\includegraphics[width=0.99\columnwidth,trim={0cm 0cm 0cm 0cm},clip]{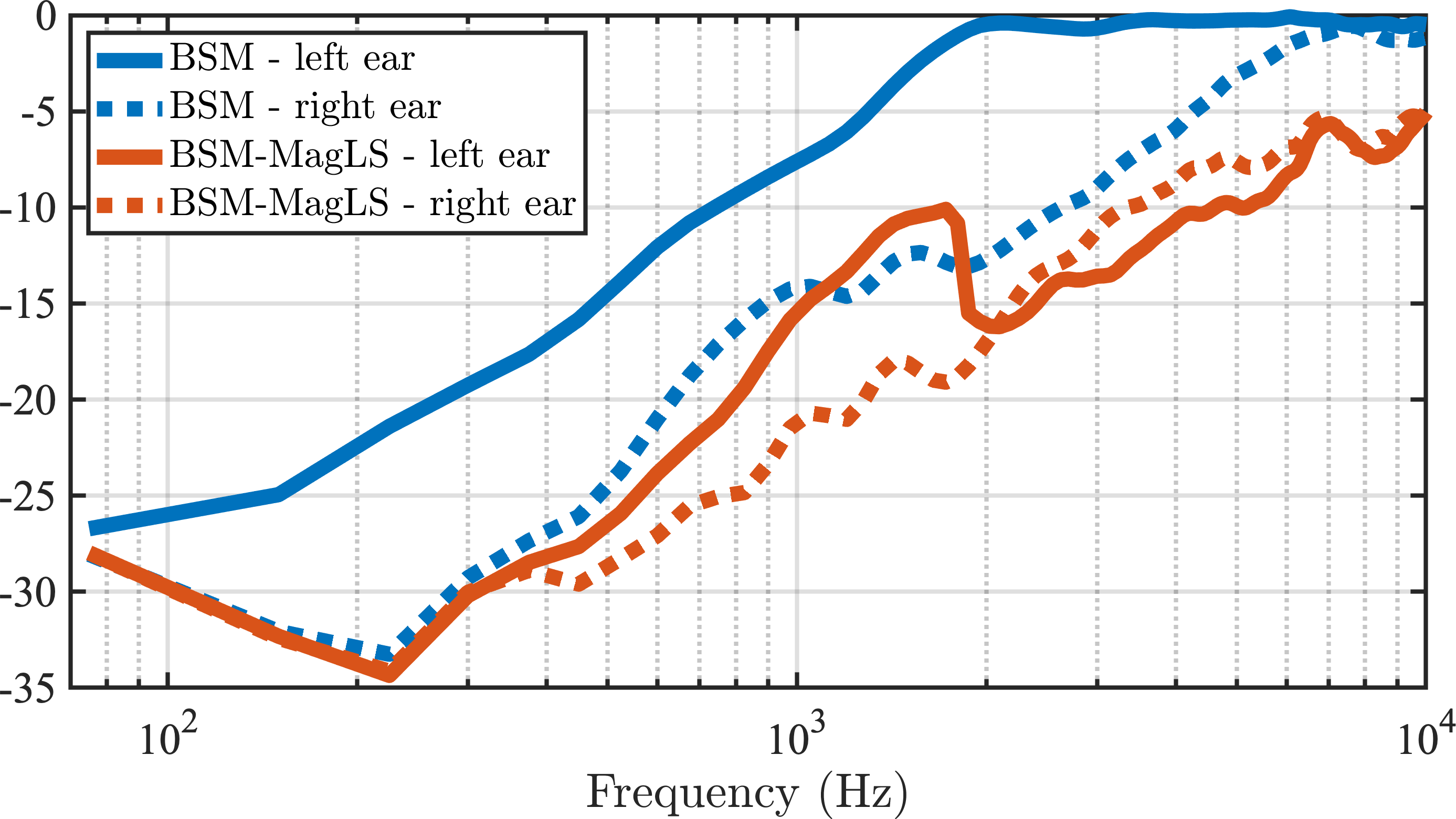}};
				\node[overlay, left=of img, node distance=0cm, rotate=90, anchor=center,yshift=-0.9cm,font=\color{black}] {\small $\bar{\epsilon}^{\,l,r}_{\text{abs}}(k)$  [dB]};
			\end{tikzpicture}	
		}
	\end{minipage}	
	\begin{minipage}{1\columnwidth}
		\centering
		\caption{Similar to Fig.~\ref{fig:eps_abs_norm_l_static} but with the BSM filters corresponding to compensation of (a) $\Delta\phi=30^{\circ}$ and (b) $\Delta\phi=60^{\circ}$ head rotations, and presented for both ears. \label{fig:eps_abs_norm_head_rot}	}		
	\end{minipage}
	
\end{figure}

Assume that the listener's head is rotated by $\Delta\phi$ degrees with respect to the reference head position during the recording phase, as described in Section~\ref{sec:head_rot}.
This is illustrated in Fig.~\ref{fig:head_array_diagrams}, where the reference head position can be seen in Fig.~\ref{fig:head_array}, and a head rotation of $\Delta\phi$ degrees is illustrated in Fig.~\ref{fig:head_rotHead}. 
In the latter case, the HRTF vector $\mathbf{h}^{l,r}$ in the BSM filters $\mathbf{c}_{\text{BSM}}^{l,r}$ and $\mathbf{c}_{\text{BSM-MagLS}}^{l,r}$ should be modified to (\ref{eq:HRTF_rot}) to enable head tracking. The magnitude errors in this case with a head rotation of $\Delta\phi=30^{\circ}$ are presented in Fig~\ref{fig:eps_abs_norm_head_rot_30}. 
First, note that the errors of the original BSM have increased significantly for the left ear, by up to approximately 10\,dB for frequencies above 1\,kHz, compared to the static reproduction conditions in Fig.~\ref{fig:eps_abs_norm_l_static}. The right ear errors have also increased, but less significantly. This may be explained by the position of the recording array relative to the ears following the head rotation, as can be seen in Fig~\ref{fig:head_rotHead}. This rotation distances the left ear from the array, while the right ear remains relatively close to the array, and thus, estimating the left ear signal is more challenging. 
In addition, note that the accuracy of BSM-MagLS also degrades following the head rotation, but overall it is much more robust to the head rotation. 
The increase in error of the original BSM due to the head rotation in this case averaged across all studied frequencies is $\approx2\,$dB and 0.4\,dB for the left and right ears, respectively. For BSM-MagLS, these errors are $\approx0.1\,$dB and 0.9\,dB for the left and right ears, respectively. 

Similarly, the errors were calculated for a head rotation of $\Delta\phi=60^{\circ}$, which is even more challenging, and are presented in Fig.~\ref{fig:eps_abs_norm_head_rot_60}. 
The original BSM produces the binaural signals with even larger errors, which are above -10\,dB for frequencies higher than 800\,Hz, and above 2\,kHz the errors are approximately 0\,dB. 
However, the BSM-MagLS errors remain relatively stable, compared to the results for $\Delta\phi=30^{\circ}$, and the errors remain below -10\,dB for frequencies below 5\,kHz. 
The increase in error of the original BSM due to the head rotation in this case averaged across all studied frequencies is $\approx2.7\,$dB and 1\,dB for the left and right ears, respectively. For BSM-MagLS, these errors are $\approx0\,$dB and 0.5\,dB for the left and right ears, respectively. 
Overall, BSM-MagLS is expected to produce the magnitude of the binaural signals much more accurately than the original BSM, when head rotations are compensated for. 

\subsection{ITD and ILD analysis}
The performance measures studied to this point are based on the MSE, which may only partially represent human perception. 
Hence, a study with perceptually-motivated measures is presented next based on ITD and ILD. 
Both measures are calculated for sound fields comprised of a single plane wave with a DOA of $(\theta,\phi)$, where $\theta=90^{\circ}$ and $\phi$ is in the range $[0^{\circ}, 359^{\circ}]$ with $1^{\circ}$ resolution. 
The simulation parameters are similar to those used in previous sections, except for the filters $\mathbf{c}_{\text{BSM-MagLS}}^{l,r}$ that were calculated with a cutoff frequency of $1.5\,$kHz, without using cross-fade between BSM and BSM-MagLS around 1.5kHz, in order to incorporate the phase of the corresponding binaural signals. 

For ITD estimation, a cross-correlation based method, which was found to be a valid perceptual measure in \cite{ITD_comparison}, is utilized. This method involves first low-pass filtering the binaural signals. A cutoff frequency of 1.5\,kHz is used for the low-pass filter (LPF), instead of the 3\,kHz LPF suggested in \cite{ITD_comparison}. This lower cutoff frequency is appropriate because BSM-MagLS intentionally disregards phase information above 1.5 kHz. The filtering was followed by the calculation of the inter-aural cross-correlation (IACC):
\begin{align}
	\text{IACC}_{p}(\tau) = \sum_{t=0}^{T-\tau-1} p^l(t+\tau)p^r(t),
	\label{eq:IACC}
\end{align}
where $p^{l,r}(t)$ are the left and right time-domain binaural signals and $T$ is the total number of time samples. Following the IACC calculation, the ITD is estimated as:
\begin{align}
	\text{ITD}(\theta,\phi) = \text{arg}\,\underset{\tau}{\text{max}\,} \{ \text{IACC}_{p}(\tau) \}.
	\label{eq:ITD_est}
\end{align}
This ITD was calculated with the corresponding head-related impulse responses (HRIRs) as the binaural signals (therefore assuming a single impulse source signal), representing the reference ITD and denoted $\text{ITD}_{\text{ref}}(\theta,\phi) $. In addition, it was calculated with the binaural signals reproduced with the BSM and its MagLS extension. The following ITD error measure was then calculated:
\begin{align}
	\epsilon_{\text{ITD}}(\theta,\phi) = | \text{ITD}(\theta,\phi) -  \text{ITD}_{\text{ref}}(\theta,\phi)|.
	\label{eq:ITD_err}
\end{align}

The ILD was estimated with the binaural signals analyzed with ERB filter bands according to \cite{ILD_measure}:
\begin{align}
	\text{ILD}(f_c, \theta,\phi) = 10\log_{10}\frac{ \sum_{f=0}^{f_{c}^{\text{max}}} |C(f, f_c)  p^{l}(f) |^2}{  \sum_{f=0}^{f_{c}^{\text{max}}} |C(f, f_c) p^{r}(f) |^2 } ,
	\label{eq:ILD_est}
\end{align}
where $C(f, f_c)$ is the ERB filter with central frequency $f_c$ evaluated at frequency $f$, and $f_{c}^{\text{max}}$ is the maximal frequency of this ERB filter. This was implemented with 29 filter bands in the range of $[50, 6000]\,$Hz, using the Auditory Toolbox \cite{Auditory_toolbox}. This range was chosen since it corresponds to speech signals, which will be used in the listening experiment in the following section. This measure was also calculated with the HRIRs as binaural signals, once again representing the reference ILD measure, denoted $\text{ILD}_{\text{ref}}(f_c, \theta,\phi) $. 
These frequency dependent ILD measures were then averaged over the ERB filter bands as:
\begin{align}
	\text{ILD}_{\text{av}}(\theta,\phi) = \frac{1}{29} \sum_{f_c} \text{ILD}(f_c,\theta,\phi).
	\label{eq:ILD_av_est}
\end{align}
Finally, the following averaged ILD error measure was calculated:
\begin{align}
    \resizebox{1\hsize}{!}{%
        $
    	\epsilon_{\text{ILD}_{\text{av}}}(\theta,\phi) =\frac{1}{29} \sum_{f_c} | \text{ILD}(f_c, \theta,\phi) - \text{ILD}_{\text{ref}}  (f_c,\theta,\phi) |.
        $
    }
	\label{eq:ILD_err_ave}
\end{align}

\begin{figure}
	\centering
	\begin{minipage}{1\columnwidth}	
		\centering
		\subfigure[No head rotation]{
			\label{fig:ITD_no_rot}	
			\begin{tikzpicture}
				\node (img)  {\includegraphics[width=0.99\columnwidth,trim={0cm 0cm 0cm 0cm},clip]{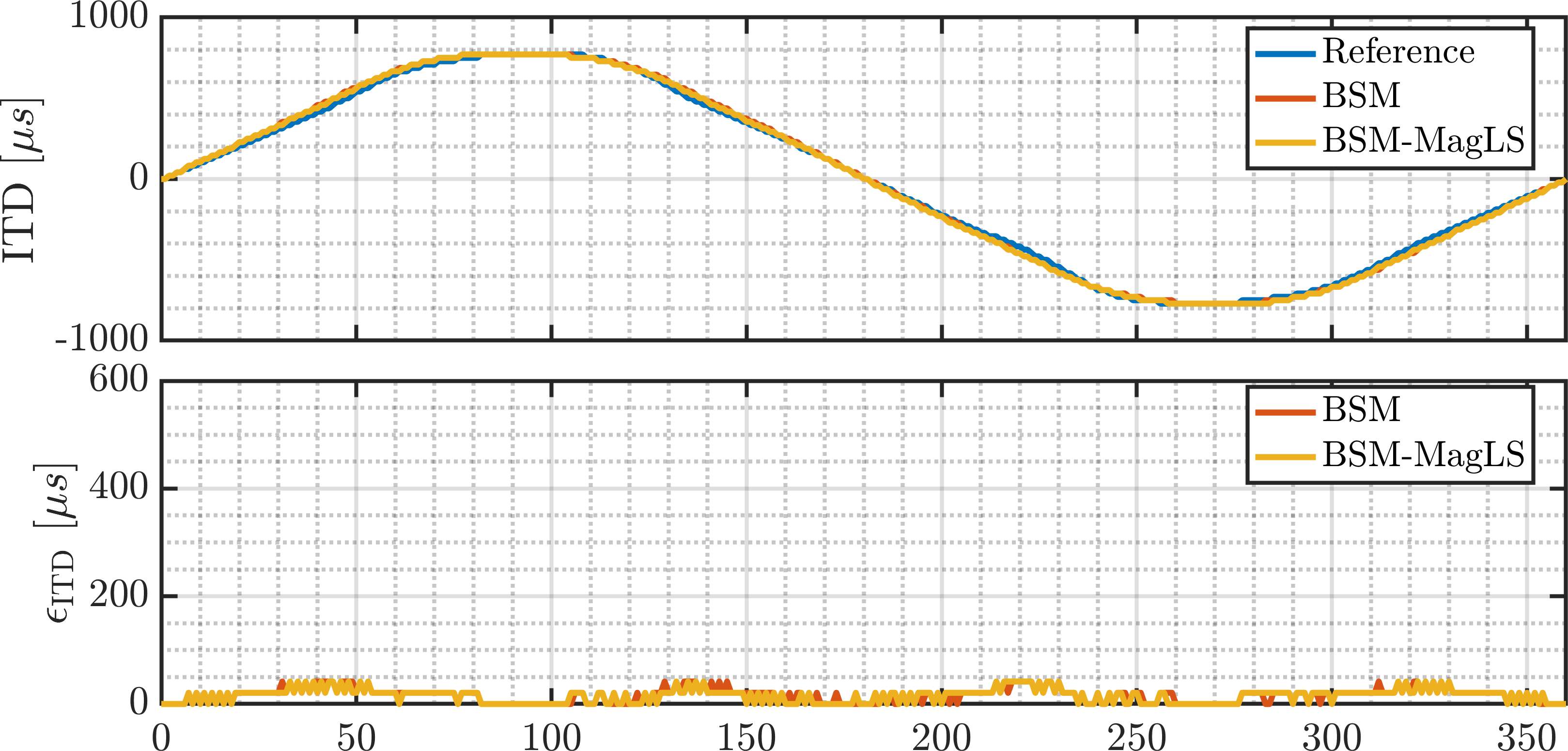}};				
				\node[overlay, node distance=0cm, rotate=0, anchor=center,yshift=-2cm,font=\color{black}] {\footnotesize Azimuth [deg]};
			\end{tikzpicture}	
		}
		~
		\centering
		\subfigure[$\Delta\phi = 30^{\circ}$]{
			\label{fig:ITD_30_rot}	
			\begin{tikzpicture}
				\node (img)  {\includegraphics[width=0.99\columnwidth,trim={0cm 0cm 0cm 0cm},clip]{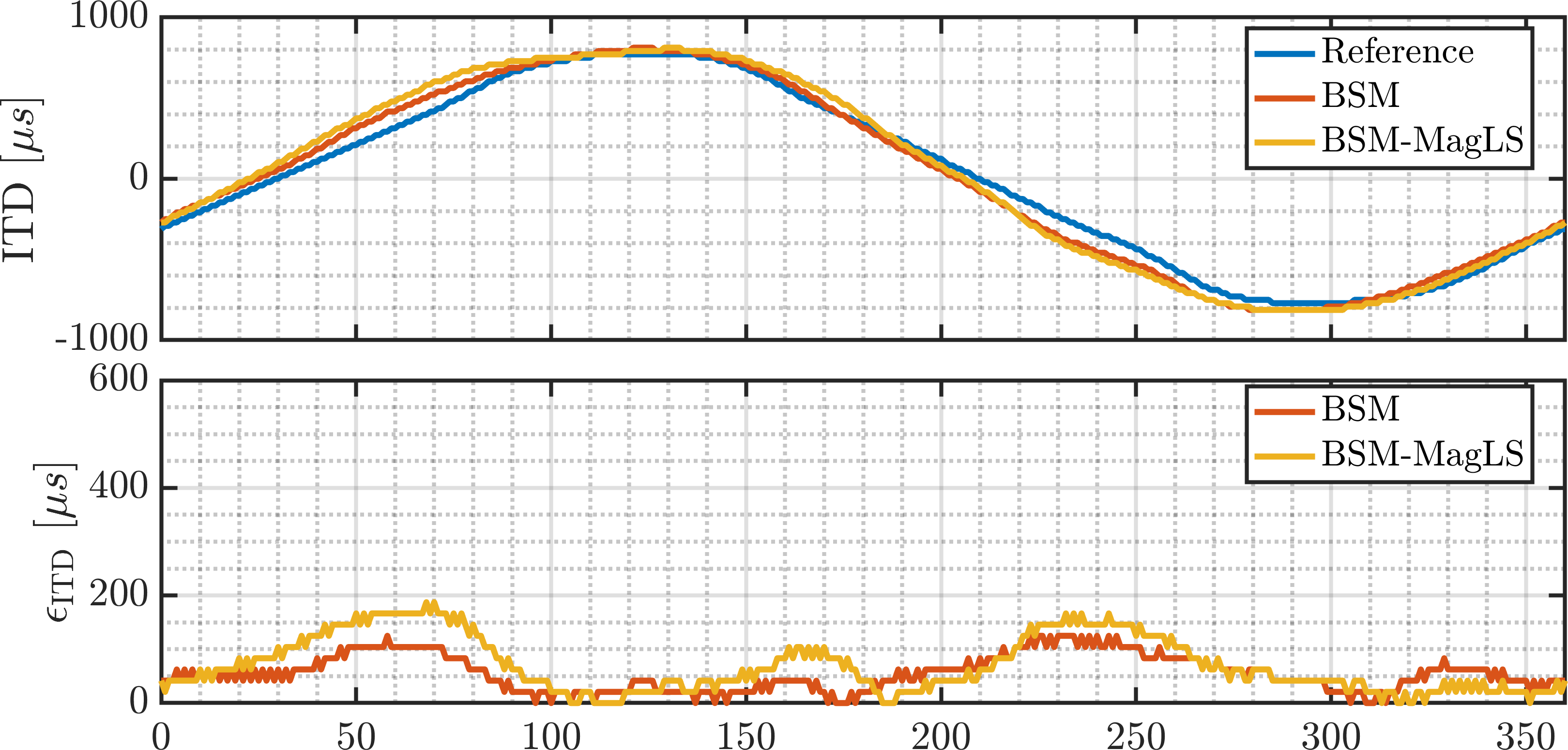}};
				\node[overlay, node distance=0cm, rotate=0, anchor=center,yshift=-2cm,font=\color{black}] {\footnotesize Azimuth [deg]};
			\end{tikzpicture}	
		}
	\centering
	\subfigure[$\Delta\phi = 60^{\circ}$]{
		\label{fig:ITD_60_rot}	
		\begin{tikzpicture}
			\node (img)  {\includegraphics[width=0.99\columnwidth,trim={0cm 0cm 0cm 0cm},clip]{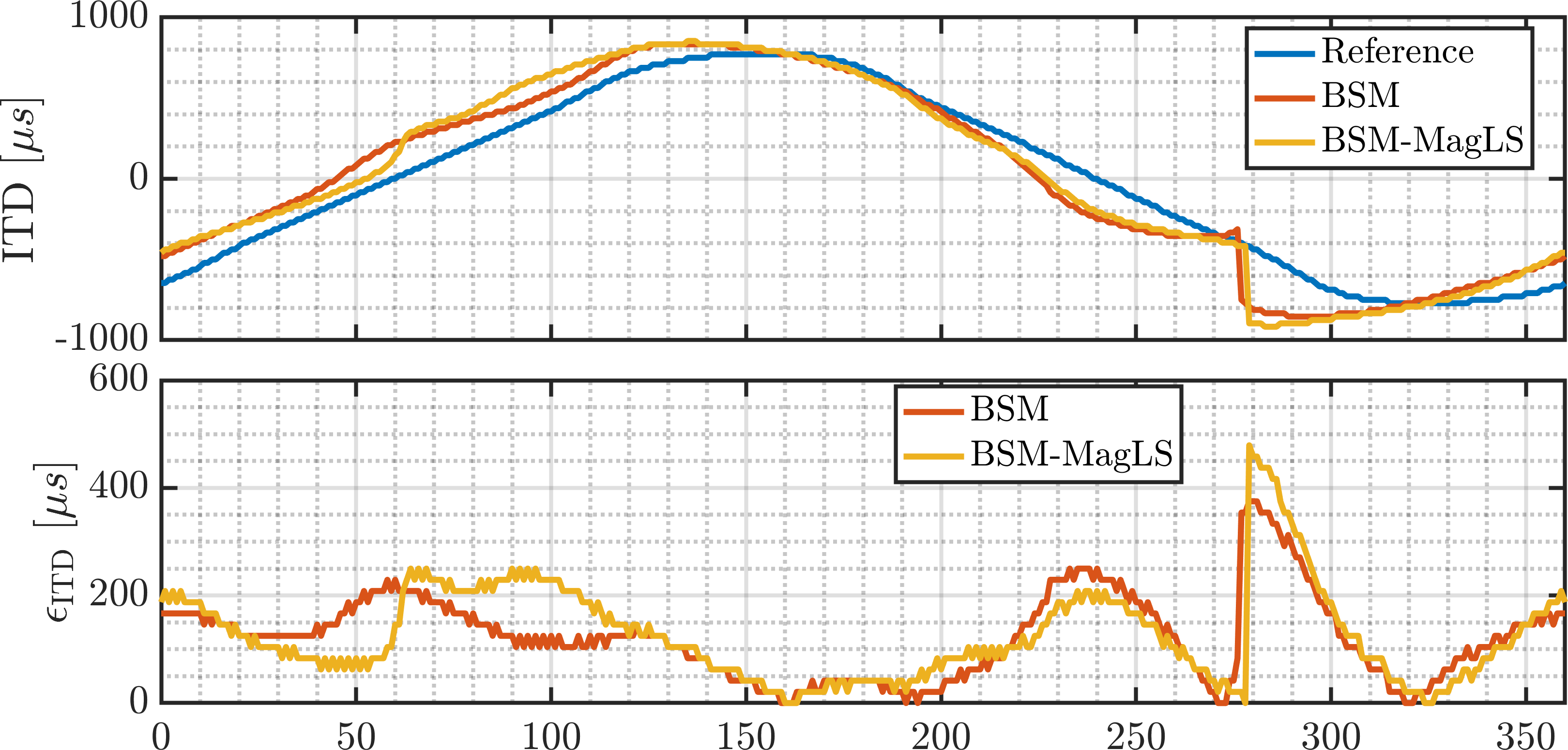}};
			\node[overlay, node distance=0cm, rotate=0, anchor=center,yshift=-2cm,font=\color{black}] {\footnotesize Azimuth [deg]};
		\end{tikzpicture}	
	}
	\end{minipage}	
	\begin{minipage}{1\columnwidth}
		\centering
		\caption{Estimated ITD (top) and ITD error (bottom) measures as in (\ref{eq:ITD_est}) and (\ref{eq:ITD_err}), respectively. The measures are calculated with the reference HRIR signal, BSM and BSM-MagLS, with a semi-circular array with $M=6$ microphones. (a) Original array orientation, (b) head rotation of $30^{\circ}$, and (c) head rotation of $60^{\circ}$. \label{fig:ITD_all}	}		
	\end{minipage}
	
\end{figure}

Figure~\ref{fig:ITD_all} presents the estimated ITD (calculated using (\ref{eq:ITD_est})) and ITD errors (calculated using (\ref{eq:ITD_err})) as a function of azimuth, obtained with the six-microphone semi-circular array. In Fig~\ref{fig:ITD_no_rot}, these measures are shown for static recording and reproduction conditions. The estimated ITDs are relatively accurate in this case, with ITD errors below the just noticeable difference (JND) threshold of $20\,\mu s$ and $100\, \mu s$ for the front and lateral directions, respectively \cite{ITD_JND_1,ITD_JND_2}. In addition, the errors are very similar between the BSM and its MagLS extension, which is expected since the ITD measures are dominated by the lower frequency range, at which both versions are designed similarly. 
The cases of compensating for head rotations during playback are presented for $\Delta\phi=30^{\circ}$ in Fig.~\ref{fig:ITD_30_rot} and for $\Delta\phi=60^{\circ}$ in Fig.~\ref{fig:ITD_60_rot}. Generally, as the degree of the rotations increases, the ITD errors increase as well. More specifically, when $\Delta\phi=30^{\circ}$, the errors for lateral angles seem to increase for up to $200\,\mu s$, and up to $400\,\mu s$ when $\Delta\phi=60^{\circ}$. However, the increase in ITD errors is an additional indication that BSM may produce binaural signals with reduced quality in terms of source localization, when compensating for head rotations. This may be overcome by using an array with higher spatial resolution, such as a fully-circular array, and by using more microphones, but the study of such arrays is out of the scope of this paper.

\begin{figure}
	\centering
	\begin{minipage}{1\columnwidth}	
		\centering
		\subfigure[No head rotation]{
			\label{fig:ILD_no_rot}	
			\begin{tikzpicture}
				\node (img)  {\includegraphics[width=0.99\columnwidth,trim={0cm 0cm 0cm 0cm},clip]{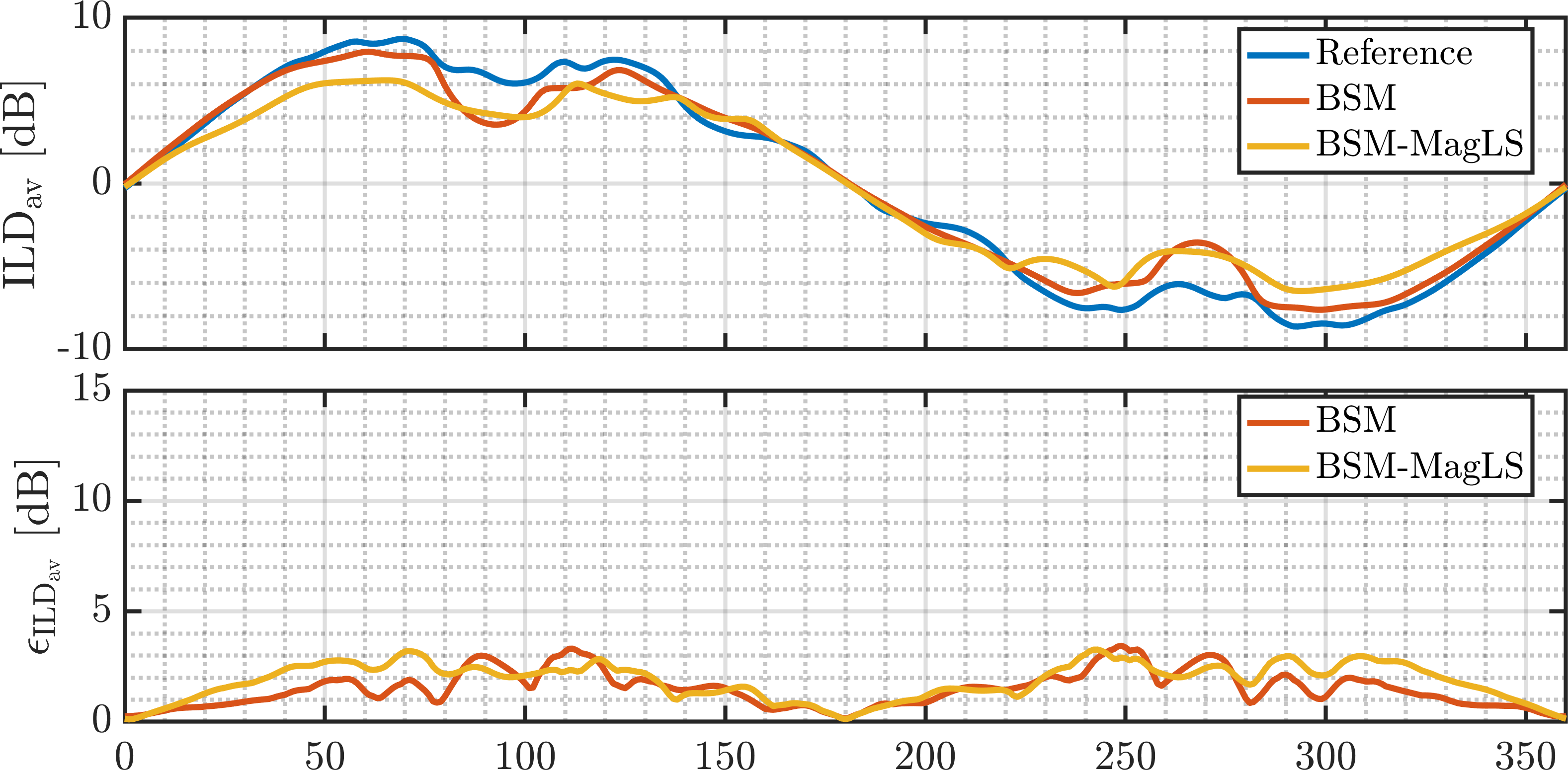}};				
				\node[overlay, node distance=0cm, rotate=0, anchor=center,yshift=-2cm,font=\color{black}] {\footnotesize Azimuth [deg]};
			\end{tikzpicture}	
		}
		~
		\centering
		\subfigure[$\Delta\phi = 30^{\circ}$]{
			\label{fig:ILD_30_rot}	
			\begin{tikzpicture}
				\node (img)  {\includegraphics[width=0.99\columnwidth,trim={0cm 0cm 0cm 0cm},clip]{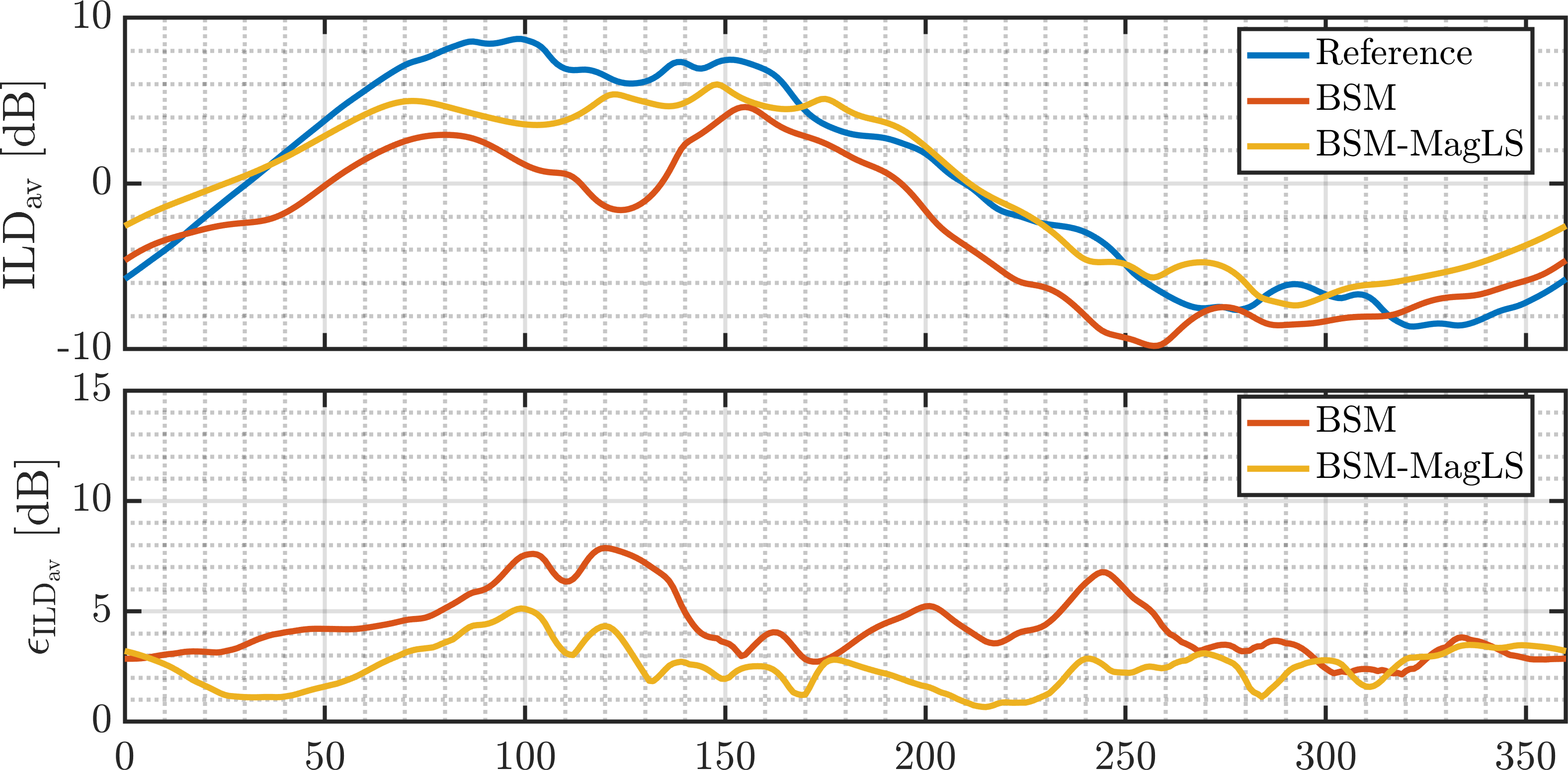}};
				\node[overlay, node distance=0cm, rotate=0, anchor=center,yshift=-2cm,font=\color{black}] {\footnotesize Azimuth [deg]};
			\end{tikzpicture}	
		}
		\centering
		\subfigure[$\Delta\phi = 60^{\circ}$]{
			\label{fig:ILD_60_rot}	
			\begin{tikzpicture}
				\node (img)  {\includegraphics[width=0.99\columnwidth,trim={0cm 0cm 0cm 0cm},clip]{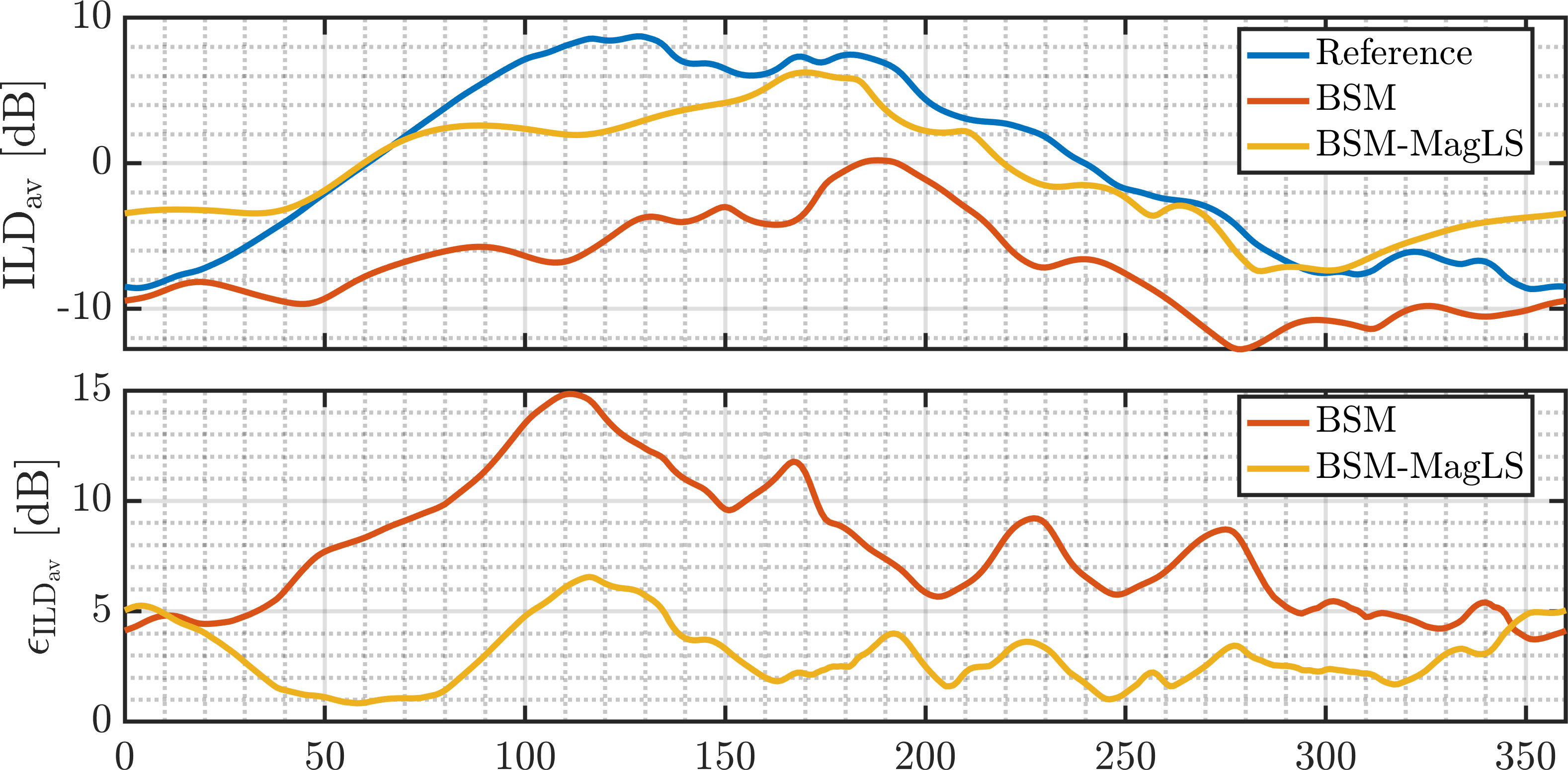}};
				\node[overlay, node distance=0cm, rotate=0, anchor=center,yshift=-2cm,font=\color{black}] {\footnotesize Azimuth [deg]};
			\end{tikzpicture}	
		}
	\end{minipage}	
	\begin{minipage}{1\columnwidth}
		\centering
		\caption{Estimated average ILD (top) and averaged ILD error (bottom) measures as in (\ref{eq:ILD_av_est}) and (\ref{eq:ILD_err_ave}), respectively. The measures are calculated with the reference HRIR signal, BSM and BSM-MagLS, with a semi-circular array with $M=6$ microphones. (a) Original array orientation, (b) head rotation of $30^{\circ}$, and (c) head rotation of $60^{\circ}$. \label{fig:ILD_all}	}		
	\end{minipage}
	
\end{figure}

The previously described ILD measures are presented in Fig.~\ref{fig:ILD_all} as a function of azimuth. The static recording and reproduction conditions which are presented in Fig.~\ref{fig:ILD_no_rot} produce relatively small ILD errors for both the BSM and its MagLS extension. However, most lateral angles seem to correspond to errors which are above the JND threshold of 1\,dB \cite{ILD_JND_1, ILD_JND_2}. The cases of compensating for head rotations of $\Delta\phi=30^{\circ}$ and $\Delta\phi=60^{\circ}$ are presented in Fig.~\ref{fig:ILD_30_rot} and Fig.~\ref{fig:ILD_60_rot}, respectively. Once again, the errors increase in both cases with the degree of rotation. However, BSM-MagLS achieves much smaller errors compared to the original BSM reproduction for almost all azimuth angles, with errors lower by up to 4\,dB when $\Delta\phi=30^{\circ}$, and up to $9\,$dB when $\Delta\phi=60^{\circ}$. 
These results demonstrate the advantage of incorporating BSM-MagLS at high frequencies for the magnitude reproduction of binaural signals.

\section{Listening Experiment}
\label{sec:list_exp}
This section presents a listening experiment that aims to subjectively analyze the quality of the BSM methods. More specifically, the original BSM and its MagLS extension will be compared when the acoustic environment is comprised of reverberant speech, using an egocentric microphone array that is mounted on a pair of glasses (different than the semi-circular array studied in Section~\ref{sec:simulations}, and including listener head rotations during playback. For this purpose, two acoustic environments with different characteristics were simulated and are described next.

\subsection{Setup}
\label{subsec:list_exp_setup}
\begin{figure}
	\centering
	\begin{minipage}{0.7\columnwidth}	
		\centering
		\begin{tikzpicture}
			\node (img)  {\includegraphics[width=0.99\columnwidth,trim={0cm 0cm 0cm 0cm},clip]{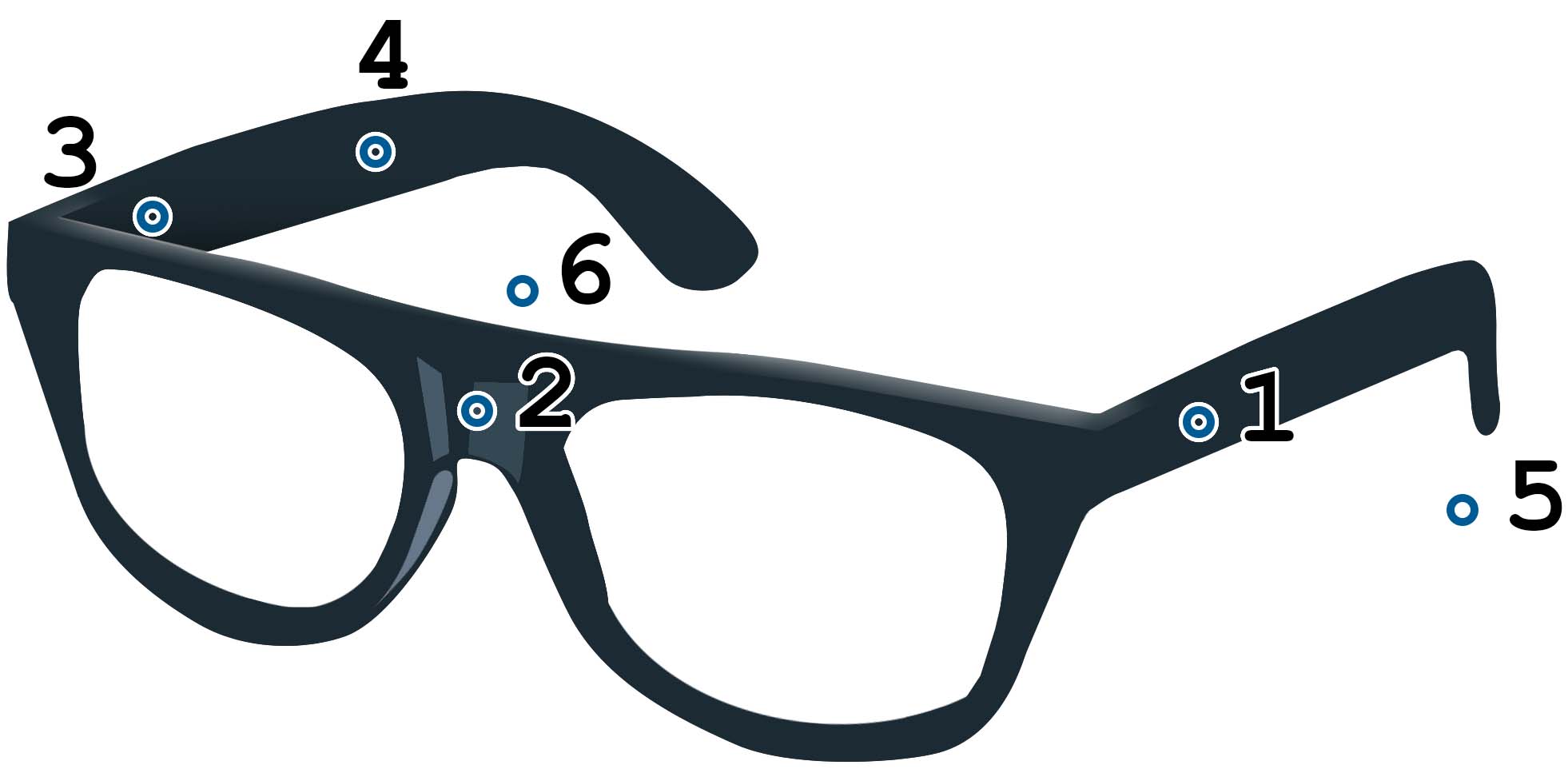}};
		\end{tikzpicture}	
	\end{minipage}	
	\begin{minipage}{1\columnwidth}
		\centering
		\caption{An illustration of the microphone array mounted on a pair of glasses that was used in the listening experiment \cite{easycom}. Only the microphones labeled 1-4 were used for performing the BSM reproduction. \label{fig:ar_glasses}}
	\end{minipage}
\end{figure}

To generate the listening experiment signals, a point source was simulated inside a shoe-box room using the Multi-Channel Room Simulator (MCRoomSim) \cite{MCRoomSim} in MATLAB \cite{MATLAB}. The point source positions, room dimensions, and reverberation time for each room are described in Table~\ref{table:1}. Note that two sets of room parameters were chosen to increase the diversity of the test conditions. 
For the source signals, four seconds long speech was used, taken from the TSP database \cite{Kabal02tspspeech}, sampled at 48\,kHz. Room \#1 contains female speech and room \#2 contains male speech. The directivity pattern of human voices was simulated using MCRoomSim as well, according to the gender of each speech signal. 
In order to create more realistic signals, a glasses-mounted microphone array is studied here, described in detail in \cite{easycom}, and with the array center position described in Table~\ref{table:1}. The array, illustrated in Fig.~\ref{fig:ar_glasses} is comprised of six-microphones (four mounted on the glasses and additional two microphones located approximately at the ears positions), but only the four glasses-mounted microphones were used here. The array steering vectors were measured on a head and torso simulator in an anechoic chamber, as described in \cite{easycom}. 
The array signals were calculated by convolving the measured array steering vectors and the reverberant sound-field using SH representation. More specifically, the reverberant sound field at the origin was calculated by convolving the SH representation of the room impulse response with the clean speech signal. Then, this reverberant sound field density function was transformed to the space domain at the discrete directions corresponding to the measured array steering vectors. Finally, these two functions were convolved for each microphone position to generate the array recordings. Throughout this process, a maximal SH order of 14 was used, leading to only minor errors for this binaural signal \cite{rafaely2010interaural, zhang2010insights}.

\begin{table*}
	\caption{Parameters used to generate the listening experiment signals.}
    \begin{tabular}{l|ll}
			& Room \#1                                  & Room \#2                                   \\ \hline
			Room dimensions {[}m{]}                      & $10 \times 6 \times 3$                  & $8 \times 5 \times 3$                    \\
			Reverberation time {[}sec{]}                 & 0.34                                    & 0.69                                     \\
			Source position {[}m{]}                      & (5, 4.5, 1.7)                           & (6, 2, 1.7)                              \\
			Array position {[}m{]}                       & (2, 2, 1.7)                             & (4, 4, 1.7)                              \\
			Source relative position $(r, \theta, \phi)$ & $(3.9\text{m}, 90^{\circ}, 40^{\circ})$ & $(2.8\text{m}, 90^{\circ}, 315^{\circ})$
    \end{tabular}
	\label{table:1}
\end{table*}

\subsection{Methodology}
\label{subsec:list_exp_meth}
Based on the generated signals, the BSM and its MagLS extension filters were calculated according to (\ref{eq:c_BSM}) and (\ref{eq:c_MLS_BSM}), respectively, similarly to as described in Section~\ref{sec:simulations}. More specifically, the MagLS solution was calculated as described in Section~\ref{subsec:sim_MLS_head_rot} and with a cutoff frequency of 1.5\,kHz. 
In addition, compensations for head rotations of the listener during playback with $\Delta\phi=30^{\circ}$ and $60^{\circ}$ were simulated. This was performed by modifying the HRTFs in the BSM filter calculations to the rotated version as described in (\ref{eq:HRTF_rot}). Similar to Section~\ref{subsec:simulation_setup}, the Neumann KU100 measured HRTFs were used.

The listening experiment performed here follows the MUltiple Stimuli with Hidden Reference and Anchor (MUSHRA) protocol \cite{MUSHRA}. 
The listening experiment was conducted in a static regime, but in order to study the quality of head-rotation compensation explicitly, the experiment includes signals that correspond to simulated head-rotation configurations at specific angles. 
Thus, for each room, three reproduction scenarios were generated, one for static binaural reproduction ($\Delta\phi=0^{\circ}$) and two for simulated head rotations with compensation - one for each rotation degree ($\Delta\phi=30^{\circ}$ and $60^{\circ}$).
In each scenario, the reference was chosen to be an HOA signal of order $N=14$, leading to only minor order-truncation errors for speech signals \cite{rafaely2010interaural,zhang2010insights}, calculated as described in (\ref{eq:binau_repr_SH_finite}). In addition, a first order Ambisonics (FOA) signal was calculated (also according to (\ref{eq:binau_repr_SH_finite}) but with $N_p=1$), and used as an anchor signal that represent a low-order reproduction.
Hence, the four test signals in each scenario are the hidden reference (HOA), the BSM, its MagLS extension, and the anchor (FOA), leading to a total of 24 test signal. All test signals were normalized to the range of [-1.0, 1.0] prior to writing the final audio files.

Each combination of room ID and degree of head rotation $\Delta\phi$ was presented in a separate screen and in a random order for each subject. 
The scoring criterion for evaluating the similarity between the test signals and the reference signal was defined as \textit{overall quality}, which was described to the subjects as both spatial and timbral variations \cite{brinkmann2022audio}. This criterion was used since some of the test signals contain significant spectral distortions, such that focusing separately on spatial and timbral attributes may be difficult. This criterion is scored in a scale of 0-100, where 100 means that the test signal is indistinguishable from the reference. 12 subjects (two females and 10 males) with no known hearing impairments participated in the experiment. The subjects performed the experiment in the same room, using the same hardware, which includes the AKG K-702 headphones with headphone compensation filters taken from \cite{CologneDatabase}. Prior to the listening test, two training stages were performed, the first for familiarizing the subject with the scoring criterion and the second for familiarization with the quality of binaural reproduction for each method, but with different signals than those used in the listening experiment at the following stage. An approval to involve human subjects in this experiment was provided by the ethics committee of Ben-Gurion University of the Negev.

\subsection{Results}
\label{subsec:list_exp_res}
\begin{figure}
	\centering
	\begin{minipage}{1\columnwidth}	
		\centering
		\begin{tikzpicture}
			\node (img)  {\includegraphics[width=0.99\columnwidth,trim={0cm 0cm 0cm 0cm},clip]{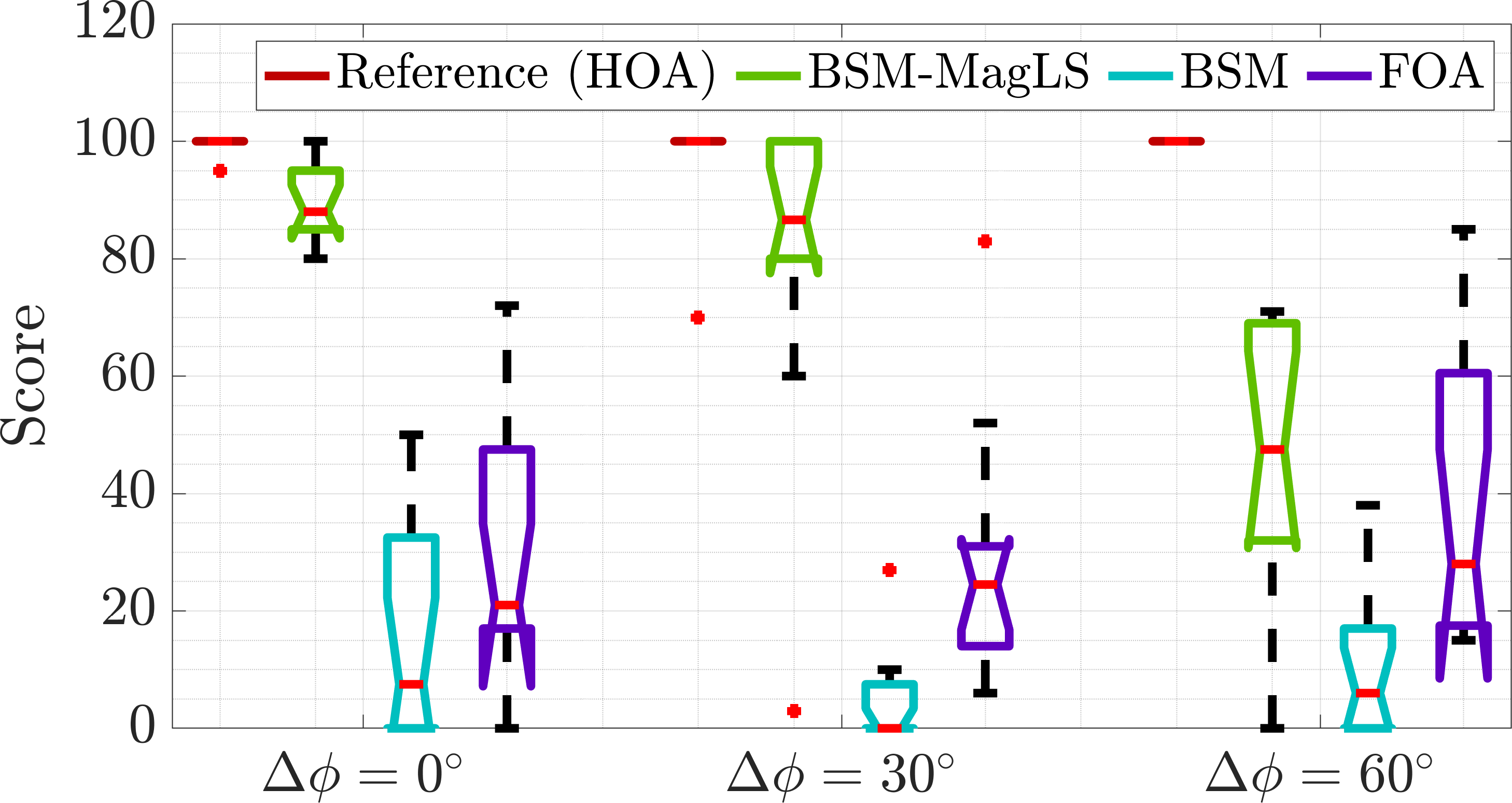}};
		\end{tikzpicture}	
	\end{minipage}	
	\begin{minipage}{1\columnwidth}
		\centering
		\caption{Box-plot of the score given by participants to each binaural reproduction method in the listening experiment, calculated over the two room IDs. The category of head rotation $\Delta\phi$ is indicated on the $x$-axis. The median is marked by the horizontal red line, outliers are marked by red plus signs, and the 25th and 75th percentiles are marked by the bottom and top colored edges, respectively. The minimal and maximal grades are marked by the black lines. Non-overlapping notches between two box-plots from the same category of head rotation indicate that the medians differ with 95\% confidence.\label{fig:mushra_means}}
	\end{minipage}
\end{figure}

The scores given by the participants to each test signal were analyzed using a repeated measures ANOVA with three within-subject factors and their interaction: (a) the room ID (\#1, \#2 as described in Table~\ref{table:1}), (b) the degree of head rotation ($\Delta\phi=0^{\circ}, 30^{\circ}, 60^{\circ}$), and (c) the binaural reproduction method (HOA, BSM-MagLS, BSM, FOA). As Mauchly's test indicated a small violation of sphericity ($\epsilon>0.75$) for most factors and interactions, and only a mild violation ($\epsilon >0.4$) for the interaction of $\Delta\phi$ - method and $\Delta\phi$ - method - room, the Greenhouse-Geisser correction was used. 
This analysis uncovered main effects for $\Delta\phi$, $F(1.69, 18.62)=13.12$, $p<.001$, $\eta_p^2=.54$, the binaural reproduction method, $F(1.89, 20.80)=154.18$, $p<.001$, $\eta_p^2=.93$, and their interaction, $F(2.97, 32.67)=9.04$, $p<.001$, $\eta_p^2=.45$, and for the room - binaural reproduction method interaction, with $F(2.28,25.13)=8.76$, $p<0.01$, $\eta^2 = 0.44$. No main effects were found for the room ID and the other interactions. 
Since the interaction between head rotation and binaural reproduction method is statistically significant, a post-hoc test with a Bonferroni correction was performed with this interaction only (excluding interaction with the room ID) and is described next. In addition, Fig.~\ref{fig:mushra_means} shows the box-plots of the scores given to each reproduction method, calculated over the two room IDs.

Next, the interaction is studied for a fixed degree of head rotation. 
In the case of static binaural reproduction ($\Delta\phi=0^{\circ}$), the mean score of the reference (HOA) was larger by 6.79 points than that of the BSM-MagLS method, and this difference was statistically significant, $p=.003$. In addition, the mean score of BSM-MagLS was larger than those of BSM and FOA by 77.29 and 66.54 points, respectively, and these difference were statistically significant, $p<.001$ in both cases. 
When compensating for a head rotation of $\Delta\phi=30^{\circ}$, the mean difference between the scores of the reference and BSM-MagLS was not statistically significant, $p=.21$. In addition, the mean scores of BSM-MagLS reproduction were larger than those of BSM and FOA by 79.3 and 54.3 points, respectively, and these differences were statistically significant, $p<.001$ for both cases. 
When comparing the mean scores of BSM and FOA reproductions for a head rotation of $\Delta\phi=30^{\circ}$, there is a differences of 25 points in favor of the FOA signal, which was statistically significant, $p=.003$. 
Finally, for the scenario of compensating a $\Delta\phi=60^{\circ}$ head rotation, the mean score of the reference was larger by 38.41 points than that of the BSM-MagLS method, and this difference was statistically significant, $p<.001$. In addition, the mean scores of BSM-MagLS reproduction were larger than those of BSM by 51.2 points, and this difference was statistically significant, $p<.001$. However, the mean score difference of BSM-MagLS and FOA was not found to be statistically significant, $p=.07$. In addition, there is a difference of 23.45 points in favor of the FOA signal over BSM, which was statistically significant, $p=.014$. 

It can be concluded that BSM-MagLS with a cutoff frequency of 1.5\,kHz may produce better binaural signals compared to the original BSM even when no head rotation compensation is required. This corresponds to the results in Fig.~\ref{fig:eps_abs_norm_l_static}, with larger errors for BSM, although they correspond to a different array than the one studied here. In addition, when head rotations are presented and should be compensated for, the BSM-MagLS method may produce binaural signals which are significantly better than those produced by the original BSM. However, as the degree of head rotation increases, and the compensated reproduction is becoming more challenging, the quality of BSM-MagLS may degrade significantly, compared to a HOA reproduction.
Finally, BSM-MagLS produces binaural signals that are comparable to the quality achieved by HOA for static reproduction conditions, while the original BSM is closer to the quality of FOA reproduction. 
While these results provide additional insights into the quality of BSM w.r.t to HOA and FOA reproduction, a more comprehensive experiment that includes other state-of-the-art binaural reproduction methods is beyond the scope of this paper and is proposed for future work.

\section{Conclusions}
\label{sec:concluions}
In this work, binaural reproduction methods designed for arbitrary microphone arrays were studied. The BSM method can produce accurate binaural signals with a six-microphone semi-circular array for frequencies lower than approximately 1.5\,kHz. This accurate reproduction can be achieved for relatively complex acoustic environments. However, the accuracy degrades significantly in the higher frequency range or when head rotation is compensated for. In these cases, BSM-MagLS can produce much more accurate binaural signals. Furthermore, using only four-microphones, this method was shown to produce binaural signals that are comparable to HOA reproduction of order $N=14$ when the degree of head rotation that is compensated for is not too large. This was shown for acoustic scenes comprised of reverberant speech, and hence, it may be very useful for teleconferencing and augmented reality applications. 

\section{Declarations}


\subsection{Funding}
This research was supported in part by Reality Labs Research at Meta and Ben-Gurion University of the Negev.



\bibliographystyle{IEEEtran}
\bibliography{myBib}

\end{document}